\documentclass[a4paper,11pt]{article}
\pdfoutput=1
\usepackage{jcappub} 
 \usepackage{ulem}
\usepackage{slashbox}
\usepackage[lofdepth,lotdepth,caption=false]{subfig}
\usepackage[T1]{fontenc}

\title{IceCube events and decaying dark matter: hints and constraints}

\author[a]{Arman Esmaili,}
\author[a]{Sin Kyu Kang}
\author[b]{and Pasquale Dario Serpico}

\affiliation[a]{Institute of Convergence Fundamental Studies, Seoul National University of Science and Technology, Gongreung-ro 232, Nowon-gu, Seoul 139-743, Korea}
\affiliation[b]{LAPTh, Univ. de Savoie, CNRS, B.P.110, Annecy-le-Vieux F-74941, France}

\emailAdd{arman@ipm.ir}
\emailAdd{neutrino.skk@gmail.com}
\emailAdd{serpico@lapth.cnrs.fr}

\subheader{LAPTH-223/14}
\abstract{In the light of the new IceCube data on the (yet unidentified) astrophysical neutrino flux in the PeV and sub-PeV range, we present an update on the status of decaying dark matter interpretation of the events. In particular, we develop further the angular distribution analysis and discuss the perspectives for diagnostics. By performing various statistical tests (maximum likelihood, Kolmogorov-Smirnov and Anderson-Darling tests) we conclude that currently the data show a mild preference (below the two sigma level) for the angular distribution expected from dark matter decay vs. the isotropic distribution foreseen for a conventional astrophysical flux of extragalactic origin. Also, we briefly develop some general considerations on heavy dark matter model building and on the compatibility of the expected energy spectrum of decay products with the IceCube data, as well as with existing bounds from gamma-rays. Alternatively, assuming that the IceCube data originate from conventional astrophysical sources, we derive bounds on both decaying and annihilating dark matter for various final states. The lower limits on heavy dark matter lifetime improve by up to an order of magnitude with respect to existing constraints, definitively making these events---even if astrophysical in origin---an important tool for astroparticle physics studies.}

\begin{document}
\maketitle
\flushbottom

\section{Introduction}
\label{sec:intro}

In the last few years, one of the most remarkable events in astroparticle physics is the discovery of a high-energy astrophysical neutrino flux by the IceCube experiment: the initial hint only consisted of a couple of PeV events~\cite{Aartsen:2013bka}; later, additional shower and track events at lower energies were unveiled~\cite{Aartsen:2013jdh}, and eventually the discovery was confirmed and strengthened in the 3 years dataset~\cite{Aartsen:2014gkd}. While the origin of these events is still under discussion, they undoubtedly serve as a powerful diagnostic tool for a number of astrophysical as well particle physics models (for some speculations see~\cite{Kistler:2013my,Murase:2013ffa,Murase:2013rfa,Anchordoqui:2013qsi,Laha:2013lka,Razzaque:2013uoa,Chen:2013dza,Ahlers:2013xia,Lunardini:2013gva,Ema:2013nda,Tamborra:2014xia,Ioka:2014kca,Ng:2014pca,Kachelriess:2014oma,Ahlers:2014ioa,Bai:2014kba,Ibe:2014pja,Bhattacharya:2014sta,Sahu:2014fua,Blum:2014ewa,Barger:2013pla} and for a review see~\cite{Anchordoqui:2013dnh}). 

In this respect, it was for instance quickly realized that the highest energy events~\cite{Feldstein:2013kka} or even the totality of the signal~\cite{Esmaili:2013gha} might be attributed to relatively heavy (PeV-scale) dark matter (DM) decay (for the follow-up works see~\cite{Bai:2013nga,Bhattacharya:2014vwa,Zavala:2014dla,Bhattacharya:2014yha,Rott:2014kfa}). For the self-consistency of the article, the key model-independent features of this putative DM signal are summarized in section~\ref{sec:fluxcomp} (for a review of decaying DM see~\cite{Ibarra:2013cra}.)
 
In the light of the extended data set of IceCube~\cite{Aartsen:2014gkd} and since the source(s) of these events have not been identified yet, we address more extensively some topics of relevance for the DM interpretation of data. First, under the assumption that DM decay is responsible for (the bulk of) the events, we study the angular distribution of the data and its compatibility with the expected angular distribution from DM decay. To this aim we make use of several statistical tests, such as maximum likelihood, Kolmogorov-Smirnov and Anderson-Darling tests, and we also discuss some limitations and possible improvements. In particular, we estimate the statistics needed to distinguish between DM-like and isotropic (extragalactic astrophysical origin) angular distributions of the events and compare it with the current diagnostic power. Our main result is that currently a DM-like angular distribution of the events is somewhat preferred over an isotropic one, albeit not significantly (with a $p$-value $\sim 10\%$). Details are reported in section~\ref{sec:angular}, which is the main section of the article.

Another important issue concerns the ``plausibility'' of this scenario. While at first sight a PeV-scale DM might appear unusual, and although alternative frameworks such as the Weakly Interacting Massive Particles (WIMPs) one remain attractive, in our opinion it is worth being cautious and open-minded to avoid the \textit{streetlight effect} bias\footnote{The \textit{streetlight effect} is named after the tendency of a key loser to look for it at easier spots, that is under the streetlights. The statement of this observational bias dates back to a parable attributed to \textit{Mulla Nasreddin}, a renowned character with subtle humor existing in the literature of several Middle Eastern and Asian countries, although sometimes with a different name.}. Given the absence of signals of new physics at any laboratory experiment, most of the current searches for DM are essentially driven by the ``traditional'' techniques available to probe them. At the very least, the observed high energy neutrino events provide one more \textit{streetlight} in the challenging quest for DM identification, opening a window on the yet poorly explored PeV scale. Although we remain agnostic on the actual particle physics scenario behind such a DM candidate, and although most of our results are generic, we note that several mechanisms exist to achieve the correct abundance of PeV-scale DM (for some models addressing this issue see~\cite{Higaki:2014dwa,Harigaya:2014waa}). In section~\ref{sec:models} we make some brief considerations on an effective field theory approach to a class of models involving heavy DM, concerning the operators leading to its decay and decay products, the associated neutrino spectrum, and some comments on possible production mechanisms.       

A further issue of interest is if the proposed candidate conflicts with other existing bounds, notably gamma-ray bounds. We devote section~\ref{sec:gamma}  to address this question and conclude that at the moment the constraints are not sufficient to exclude the DM interpretation: neutrino data remain the main handle to probe the PeV-scale DM framework.

Finally, in section~\ref{sec:lifetime} we elaborate on what we can learn from IceCube events about the heavy DM properties if they will be eventually attributed to astrophysical sources. Interestingly, even in this case the IceCube events can be used to set stringent limits on the lifetime of decaying heavy DM. We derive bounds in the mass-lifetime plane for various possible standard model decay products, which are the best bounds available in a number of cases and in all cases quite competitive with gamma-ray constraints. Section~\ref{sec:conc} summarizes our conclusions and discussions of the results. Furthermore, although we focus in this paper mainly on decaying DM, in appendix~\ref{sec:ann} we also derive bounds on the annihilation cross section $\langle\sigma v\rangle$ of annihilating DM with mass below $\sim100$~TeV. This is possible since the lower threshold of IceCube data is $\sim 20$~TeV and hence the lower energy part of events can be used to constrain WIMP-like DM annihilation cross section.

\section{Neutrino flux from decaying dark matter}
\label{sec:fluxcomp}

In this section we summarize the energy and angular distributions of the expected neutrino flux from decaying DM, which closely follows the notation and terminology used in~\cite{Esmaili:2013gha}. The neutrino flux from DM decay has both a Galactic and an extragalactic contributions. While in the case of annihilating DM the latter is often subleading and dependent on the poorly known small scale clustering properties, for decaying DM the contributions of both fluxes are comparable and their relative contribution is robustly known. Some differences exist, however, in the energy and angular distributions of these components.

For the angular distribution analysis of IceCube events we use Galactic coordinates $(b,l)$, with $b$ and $l$ denoting the Galactic latitude and longitude, respectively (varying in the range $-\pi/2\leq b\leq\pi/2$ and $-\pi\leq l <\pi$). The Galactic component of neutrino flux originates from the decay of DM particles in the Milky Way halo with the following differential flux:
\begin{equation}\label{eq:halo-flux}
\frac{{\rm d} J_{\rm h}}{{\rm d}E_\nu}\left(l,b\right) = \frac{1}{4\pi\,m_{\rm DM}\,\tau_{\rm DM}} \frac{{\rm d}N_\nu}{{\rm d}E_\nu} \int_0^\infty {\rm d}s\; \rho_{\rm h}\left[r\left(s,l,b\right)\right] \;,
\end{equation}
where $m_{\rm DM}$ and $\tau_{\rm DM}$ are the DM mass and lifetime, respectively, and $\rho_{\rm h}(r)$ is the density profile of DM particles in our Galaxy as a function of distance from the Galactic center (GC), $r$. The $dN_\nu/dE_\nu$ factor is the energy spectrum of neutrinos produced in the decay of a DM particle. The dependence of Galactic neutrino flux on the Galactic coordinates originates from the off-center position of the Sun with respect to the center of DM halo, which is located at the GC. The neutrino flux at Earth is given by the line-of-sight integration over the parameter $s$, which is related to $r$ by
\begin{equation}\label{galcoord}
r(s,l,b) = \sqrt{s^2+R^2_\odot-2 s R_\odot \cos b\cos l}\,,
\end{equation}
where $R_\odot\simeq 8.5\,{\rm kpc}$ is the Sun-GC distance. 

The extragalactic component of the neutrino flux, which originates from the DM decay at cosmological distances, is isotropic to the leading order and thus independent of the Galactic coordinates. The differential flux with respect to the neutrino energy measured at Earth, $E_\nu$, is given by
\begin{equation}\label{eqn:NuFluxEG}
\frac{{\rm d}J_{\rm eg}}{{\rm d}E_\nu} = \frac{\Omega_{\rm DM}\rho_{\rm c}}{4\pi m_{\rm DM} \tau_{\rm DM}} \int_0^\infty {\rm d}z\, \frac{1}{H(z)} \;\frac{{\rm d}N_\nu}{{\rm d}E_\nu}\left[(1+z)E_\nu\right],
\end{equation}
where $H(z)=H_0 \sqrt{\Omega_\Lambda+\Omega_{\rm m}(1+z)^3}$ is the Hubble expansion rate as a function of redshift $z$ and $\rho_{\rm c}=5.5\times10^{-6}\,{\rm GeV}\, {\rm cm}^{-3}$ is the critical density of the Universe. For the cosmological parameters we take the values derived from the Planck temperature map data~\cite{Ade:2013zuv}: $\Omega_\Lambda=0.6825$, $\Omega_{\rm m}=0.3175$, $\Omega_{\rm DM}=0.2685$ and $h\equiv H_0/100\,{\rm km}\,{\rm s}^{-1}\,{\rm Mpc}^{-1}=0.6711$.

Apart from a relative normalization factor of order unity, the spectral shapes of the Galactic and extragalactic contributions are typically similar (see for instance Fig.~1 in~\cite{Esmaili:2013gha}). To a first approximation, this means that the angular and energy dependences are basically factorized. In the following we shall perform the analyses of the two observables independently; {\it i.e.}, by integrating over energy we obtain the angular distribution for the angular analysis of data and by integration over the solid angle we derive the energy distribution for the energy analysis of data. An ideal analysis should include both the energy and angular distributions simultaneously; but performing this analysis requires detailed information of the detector (such as the declination dependence of effective area, etc.) which is anyway unavailable now.   

\begin{figure}[t!]
\centering
\subfloat[PDF of DM decay]{
\includegraphics[width=0.5\textwidth]{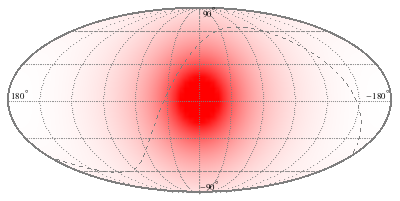}
\label{fig:pdf-dm}
}
\subfloat[PDF of IceCube data]{
\includegraphics[width=0.5\textwidth]{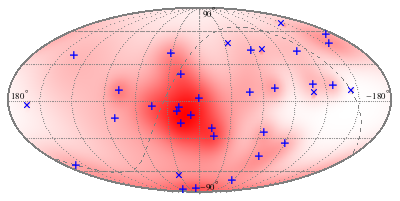}
\label{fig:pdf-data}
}
\caption{\label{fig:pdf} Illustrative sky map of: (a) DM PDF from eq.~(\ref{eqn:pDMunn2}); (b) PDF of the IceCube data from eq.~(\ref{eq:pi}). Both figures are in Galactic coordinates. The $\times$ and $+$ in panel (b) correspond to track and shower events, respectively. The color codes (linear scale) in both panels are scaled for illustration purpose.}
\end{figure}

The theoretical angular probability distribution function (PDF) from decaying DM is given by
\begin{equation}\label{eqn:pDMunn2}
p^{\rm DM}(b,l) = \kappa\left(\int_0^\infty {\rm d}s\; \rho_{\rm h}[r(s,l,b)] + \Omega_{\rm DM}\rho_c \beta\right)~,
\end{equation}
where
\begin{equation}\label{eq:beta}
\beta=\int_0^\infty \frac{{\rm d}z}{(1+z)H(z)}~,
\end{equation}
and, from the normalization condition, one has
\begin{equation}\label{eqn:pDMnorm}
\kappa=\frac{1}{4\pi \left(\eta+\Omega_{\rm DM}\rho_c\beta\right)}~,
\end{equation}
where
\begin{equation}\label{eqn:eta}
\eta= \frac{1}{4\pi}\int_{-\pi}^{\pi}  {\rm d} l \int_{-\pi/2}^{\pi/2}  {\rm d} b \cos b \int_0^\infty {\rm d}s\; \rho_{\rm h}[r(s,l,b)]\,.
\end{equation}
For our numerical analysis we adopt a Navarro-Frenk-White density profile~\cite{Navarro:1996gj}
\begin{equation}
\rho_{\rm h}(r)\simeq \frac{\rho_h}{r/r_c (1+r/r_c)^2}\;,
\end{equation}
where $r_c\simeq20\,\text{kpc}$ is the critical radius and $\rho_h=0.33\,{\rm GeV}\, {\rm cm}^{-3}$, which yields a DM density at the Solar System $\rho_\odot=0.39\,{\rm GeV}\, {\rm cm}^{-3}$~\cite{Catena:2009mf}. Figure~\ref{fig:pdf-dm} shows the illustrative PDF of decaying DM in Galactic coordinates, where the stronger reddish color depicts higher value of PDF (the color code is in linear scale). It is worth to emphasize that in the case of decaying DM, the relative contribution of Galactic (which is centered at GC) and extragalactic (which is isotropic) components is fixed; the total PDF has a mild concentration around GC and flattens out to quasi-isotropic when moving to larger latitudes and longitudes. For the case of isotropic distribution (which is the case for astrophysical neutrinos), obviously the PDF is
\begin{equation}\label{eq:isopdf}
p^{\rm iso}(b,l)=\frac{1}{4\pi}\,.
\end{equation}
In section~\ref{sec:angular} we will use the PDFs in eqs.~(\ref{eqn:pDMunn2}) and (\ref{eq:isopdf}) for the angular analysis of IceCube data.  

The expected energy spectrum of neutrinos flux from DM decay can be obtained by integrating eqs.~(\ref{eq:halo-flux}) and (\ref{eqn:NuFluxEG}) over solid angle. The average Galactic flux over the full sky can be written as
\begin{equation}\label{eq:halo}
\frac{{\rm d}J_{\rm h}}{{\rm d}E_\nu}=  D_{\rm h} \; \frac{{\rm d}N_\nu}{{\rm d}E_\nu}~,
\end{equation}
where
\begin{equation}\label{eq:Dhalo}
D_{\rm h}=1.7\times 10^{-12} 
\left(\frac{1\, {\rm PeV}}{m_{\rm DM}}\right)
\left(\frac{10^{27}\,{\rm s}}{\tau_{\rm DM}}\right)\,({\rm cm}^{2}\,{\rm s}\,{\rm sr})^{-1}\,.
\end{equation}

Similarly to eq.~(\ref{eq:halo}), for the extragalactic component one can write
\begin{equation}\label{eqn:NuFluxEG2}
\frac{{\rm d}J_{\rm eg}}{{\rm d}E_\nu} =D_{\rm eg}  \int_0^\infty \, \frac{{\rm d}z}{\sqrt{\Omega_\Lambda+\Omega_{\rm m}(1+z)^3}} \frac{{\rm d}N_\nu}{{\rm d}E_\nu}\left[\left(1+z\right)E_\nu\right]\,,
\end{equation}
where
\begin{equation}\label{eq:Deg}
D_{\rm eg}=1.4\times 10^{-12} 
\left(\frac{1\, {\rm PeV}}{m_{\rm DM}}\right) \left(\frac{10^{27}\,{\rm s}}{\tau_{\rm DM}}\right)\,({\rm cm}^{2}\,{\rm s}\,{\rm sr})^{-1}\,.
\end{equation}
Comparing eqs.~(\ref{eq:Dhalo}) and (\ref{eq:Deg}) reveals that the magnitudes of Galactic and extragalactic components are similar. In section~\ref{sec:models} the energy spectra in eqs.~(\ref{eq:halo}) and (\ref{eqn:NuFluxEG2}) will be used for computing the energy distribution of events from DM decay. 

\section{Angular distribution of IceCube data: dark matter vs. isotropy}\label{sec:angular}

In~\cite{Aartsen:2014gkd}, it is merely mentioned that the collected high energy data by IceCube have been checked to be compatible with an isotropic distribution. However, the relevant question in testing the DM scenario is whether the angular distribution of data prefer a DM-like distribution or isotropic one, and what is the significance of preference. In this section we elaborate on this question. Needless to say, a rigorous evaluation of the compatibility with (or preference for) a decaying DM distribution can only be performed within the IceCube collaboration. We show in the following that a number of tests are possible with the currently available information, under different simplifying approximations. Our preliminary conclusions are the following: Either a likelihood method, see section~\ref{likeli}, or a Kolmorogov-Smirnov test, see section~\ref{KStest} (including its Anderson-Darling variant in section~\ref{ADtest}), suggest that data prefer a DM-like distribution with respect to an isotropic one at the confidence level ranging from 89\% to 98\% C.L.; {\it i.e.}, roughly at the ``$2\,\sigma$'' level. In section~\ref{otherissues} we discuss some of the possible limitations of our treatments. We argue that, while an improved treatment of the angular resolution or background-signal separation might degrade somewhat these figures, some preference for a DM-like distribution (i.e. for some excess toward the inner Galaxy) persists. We also perform a KS forecast analysis of the number of events needed to reject the DM distribution, if the distribution were isotropic. This forecast does suggest that current diagnostic power of a KS test is not better than 1 to 2 $\sigma$ level, and that a statistics of about $120$ events is needed to reach a $\sim99\%$ C.L.. This appears within reach of the lifetime of the IceCube detector, which suggests that more conclusive tests should be within reach even with the current experiment. Definitely, a better diagnostic power can be attained by the planned next generation of neutrino detectors.

\subsection{Likelihood analysis}\label{likeli}

A first test consists in checking how well the event distribution follows a DM-like distribution as opposed to an isotropic one, via a likelihood test.
For each event we can define a probability distribution function $p_i$. We follow here the same ``flat sky'' approximation common in neutrino telescopes for
point source analyses~\cite{Braun:2008bg}, namely
\begin{equation}\label{eq:pi}
p_i (b,l) =  \frac{1}{2\pi\sigma_i^2} \exp \left[ -\frac{|\vec{x}_i-\vec{x}|^2}{2\sigma_i^2} \right],
\end{equation}     
where $|\vec{x}_i-\vec{x}|$ is the angular distance between two points and $\sigma_i$ is the error in the reconstruction of direction reported by IceCube. In this paper we analyze 35 events collected during three years at IceCube detector\footnote{IceCube observed 37 events in the search region. But two of these events have characteristics so similar to background events that they are excluded from the analysis.}~\cite{Aartsen:2014gkd}. Out of this 35 events, 7 are track events (with small uncertainty in the reconstruction of incoming neutrino direction, $\sigma\lesssim1^\circ -2^\circ$) and the remainder are shower events\footnote{\label{foot}Note that this ratio of topologies cannot be translated directly into a flavor ratio, as sometimes done in recent literature, without appropriate corrections for the cuts and experimental vetoes, which require the detector's Monte Carlo. Several of the ``shower'' events might be $\nu_\mu$ charged current interactions dominated by the shower at the interaction vertex. In some cases, the muon track could escape the detector without detection and the event will then be classified as shower. One can easily visualize this by looking at event 8 of~\cite{Aartsen:2013jdh}---which is classified as a track---and imagining moving the vertex of the event closer to the lower border of the fiducial volume, which would have caused the muon track to be missed. This adds a large uncertainty to the flavor content of the track vs shower channels since this ``topology efficiency correction'' is not currently applied. A similar correction has been mentioned in~\cite{Aartsen:2014yta}, although for a different set of data. This would also correspond to the ``trivial'' solution to the puzzle among the ones mentioned in~\cite{Mena:2014sja}.}. In Table~I in the supplementary section of~\cite{Aartsen:2014gkd}, all the events properties, including the coordinates $\vec{x}_i$ and uncertainties $\sigma_i$, are listed. In the Galactic coordinates, the angular distance writes
\begin{equation}
|\vec{x}_i-\vec{x}|^2 = \left( \arccos \left[\sin b \sin b_i + \cos b \cos b_i \cos (l-l_i) \right] \right)^2~,
\end{equation}
where $(b_i,l_i)$ is the Galactic coordinate of the $i$-th event\footnote{To check our input and basic manipulation/likelihood routines, as a preliminary test we performed the same analysis done by IceCube collaboration, that is searching for clustering of events. We could reproduce precisely the results in SUPPL. FIG. 2 and FIG. 5 in~\cite{Aartsen:2014gkd}.}. Figure~\ref{fig:pdf-data} shows a schematic view of the sky-map observed by IceCube.
 
\begin{figure}[t!]
\centering
\subfloat[]{
\includegraphics[width=0.5\textwidth]{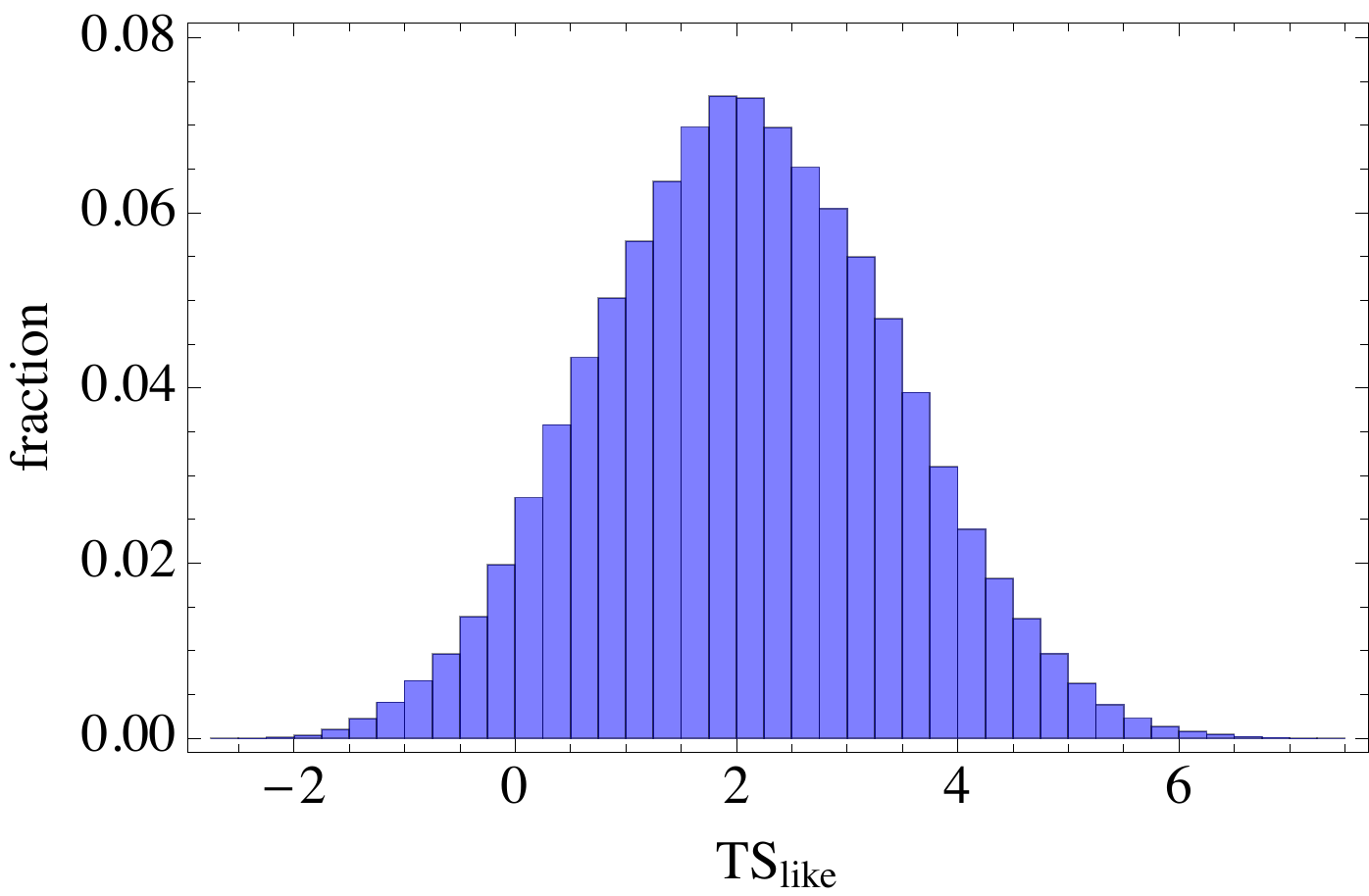}
\label{fig:LT-dis}
}
\subfloat[]{
\includegraphics[width=0.5\textwidth]{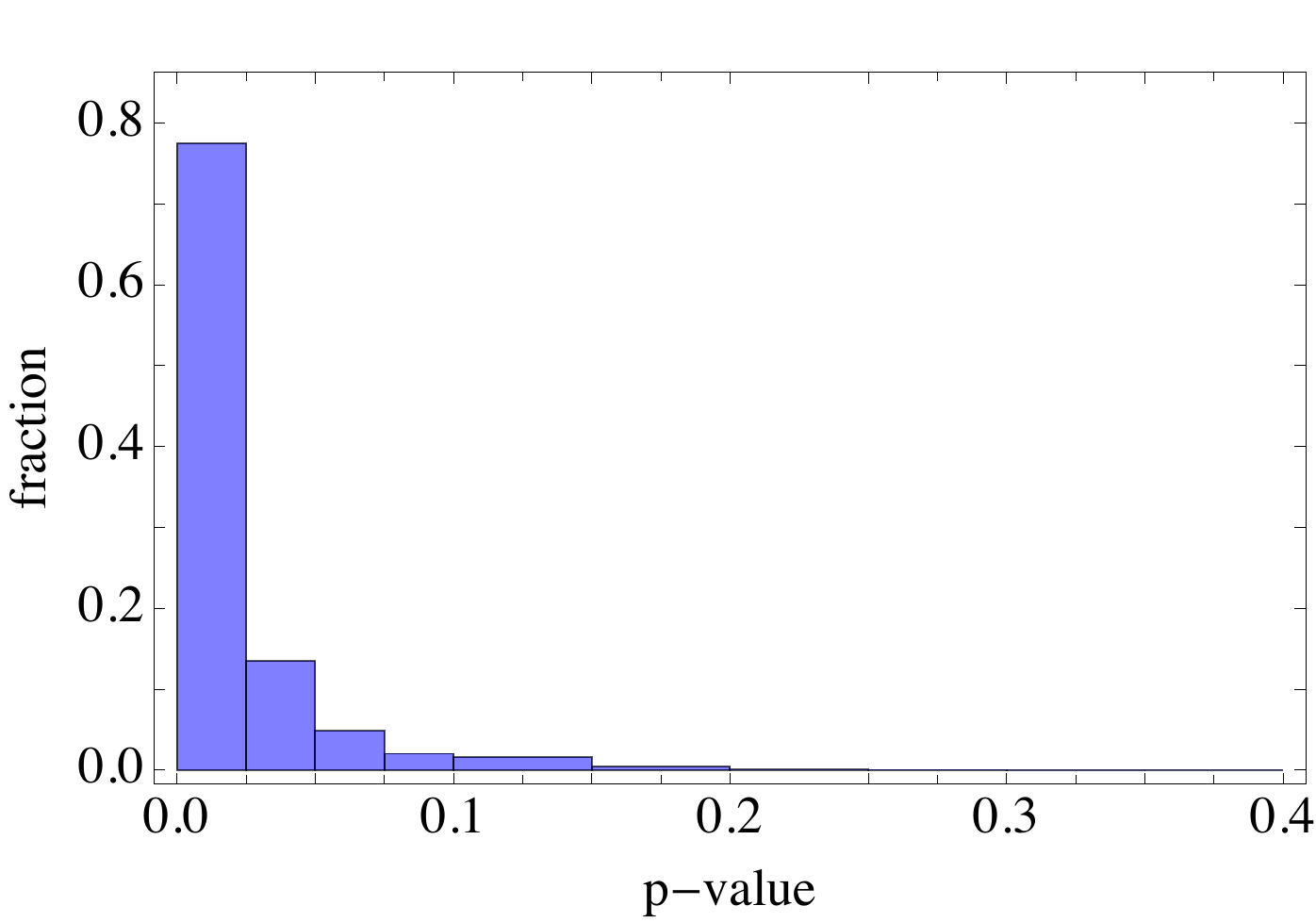}
\label{fig:LT-pvalue}
}
\caption{\label{fig:lt} Panel (a): histogram of the fraction of realizations out of $26 \choose 15$ having ${\rm TS}_{\rm like}$ value within the shown bins. Panel (b): histogram of the fraction of realizations out of $26 \choose 15$ having $p$-values in the shown bins.}
\end{figure}

The following test statistics (TS) can be defined for the likelihood analysis:
\begin{equation}\label{eq:ts-like}
{\rm TS_{like}} = 2\sum_{i=1}^N \left( \ln f_i - \ln p^{\rm iso}_{i} \right)=2 \ln \left( \prod_{i=1}^N f_i \right) -2 N \ln \left( \frac{1}{4\pi} \right)\,,
\end{equation}
where
\begin{equation}\label{eq:fi}
f_i = \int p_i (b,l)\; p^{\rm DM} (b,l) \cos(b)\; {\rm d}b\;{\rm d}l = \frac{1}{2\pi\sigma_i^2} \int  e^{-\frac{|\vec{x}_i-\vec{x}|^2}{2\sigma_i^2}} p^{\rm DM} (b,l) \cos(b)\; {\rm d}b\;{\rm d}l~.
\end{equation}
In the formulae above, $N$ refers to the number of signal events. It would be by far too optimistic to assume that $N=35$ events, since according to IceCube, a number of background events given by $N_b=15^{+10.1}_{-5.8}$ contaminates the sample. For illustrative purposes, let us assume that $N_b=15$, namely the reported central value by IceCube, which means that $N=20$ in eq.~(\ref{eq:ts-like}). However, still we do not know {\it which} events are background and which ones are signal. To simplify a bit the combinatorics problem, we can nonetheless make the assumption that none of the events with energy $E_\nu>150$~TeV is due to background (that is atmospheric neutrinos or muons) since IceCube collaboration estimates the number of background events in that range of energy $\ll 1$. This leaves ${26 \choose 15}=7,726,160$ ways of selecting which events are background, among the low energy events. For each case we calculate the ${\rm TS_{like}}$ value, whose distribution (with the mean value $\overline{{\rm TS}}_{\rm like}=2.1$) is shown in Figure~\ref{fig:LT-dis}. To estimate the $p$-value, we generated a similar distribution of ${\rm TS}_{\rm like}$ values from $\sim 10^5$ sets of isotropically distributed 20 events datasets. For each realization of ${26 \choose 15}$ set, we calculate the $p$-value by comparing its ${\rm TS}_{\rm like}$ with the ${\rm TS}_{\rm like}$ values of generated events (simply, the $p$-value is the fraction of generated events which have smaller ${\rm TS}_{\rm like}$ values than the one computed by observed data). Figure~\ref{fig:LT-pvalue} shows the distribution of computed $p$-values. It is clear graphically that data prefer mildly the DM angular distribution rather than isotropic distribution. The average $p$-value is found to be $\sim 2\%$, which means a $\sim98\%$ C.L. preference for DM distribution.   

\subsection{Kolmogorov-Smirnov test}\label{KStest}

An alternative statistical test, the KS test, can be performed to quantify the compatibility of data with DM vs. isotropic distributions. The KS test is a powerful test with several advantages, including its \textit{non-parametric} nature which means that it is independent of the underlying (unknown) distribution function of data. Although the two dimensional KS tests suffer in general of some ambiguities (such as the choice of integration region), a one-dimensional KS test is possible if one takes into account the symmetry of the problem: the DM distribution only depends on the angle $\vartheta$ measuring the angular distance from GC (i.e. the half-angle of the cone around GC with the axis passing through the GC), and not on the azimuthal angle $\varphi$ around the cone. This allows one to use $\vartheta$ as the only variable, suitable for a one-parameter KS test. The DM and isotropic distribution functions, in terms of $\vartheta$ parameter, are the followings: 
\begin{enumerate}
\item The isotropic distribution $p^{\rm iso}(\vartheta)$ is given by 
\begin{equation}
p^{\rm iso} (\vartheta) = \int_{0}^{2\pi} p^{\rm iso} (\vartheta,\varphi)\, {\rm d}\varphi = \int_{0}^{2\pi} \frac{1}{4\pi}\, {\rm d}\varphi = \frac{1}{2}~.
\end{equation}
\item The DM distribution $p^{\rm DM}(\vartheta)$ can be obtained by changing the variables from Galactic coordinates $(b,l)$ to $(\vartheta,\varphi)$, such that $\cos\vartheta = \cos b \cos l$; and it takes the following form:
\begin{equation}
p^{\rm DM} (\vartheta) = \int_{0}^{2\pi} p^{\rm DM} (\vartheta,\varphi)\, {\rm d}\varphi = \frac{\int_{0}^{\infty} \rho[r(s,\vartheta)] {\rm d}s + \Omega_{\rm DM} \rho_c\beta}{2(\eta + \Omega_{\rm DM} \rho_c\beta)}~,
\end{equation}
where $\beta$ and $\eta$ are given respectively in eqs.~(\ref{eq:beta}) and (\ref{eqn:eta}); and $r$, given in eq.~(\ref{galcoord}), takes the following form: $r(s,\vartheta)=\sqrt{s^2 + R_\odot^2 - 2sR_\odot\cos\vartheta}$~.
\end{enumerate}
Notice that for both the above PDFs, we have the normalization $\int_0^\pi p(\vartheta) \sin\vartheta\, d\vartheta =1$. The KS test compares the \textit{empirical distribution function} (EDF) of data with the cumulative distribution function (CDF) of the distribution being tested. The EDF of data is given by
\begin{equation}
{\rm EDF}^{\rm data} (\vartheta) = \frac{1}{N} \sum_{i=1}^{N} \Theta (\vartheta - \vartheta_i)
\end{equation}  
where $N$ is the number of signal events and $\Theta$ is the Heaviside step function. The CDF of DM and isotropic distributions can be calculated as:
\begin{equation}
{\rm CDF}^{\rm DM} (\vartheta) = \int_{0}^{\vartheta} p^{\rm DM} (\vartheta^\prime) \sin\vartheta^\prime \, d\vartheta^\prime~,
\end{equation}
and,
\begin{equation}
{\rm CDF}^{\rm iso} (\vartheta) = \int_{0}^{\vartheta} p^{\rm iso} (\vartheta^\prime) \sin\vartheta^\prime \, d\vartheta^\prime=\frac{1-\cos\vartheta}{2}~. 
\end{equation}

\begin{figure}[t!]
\centering
\includegraphics[width=0.6\textwidth]{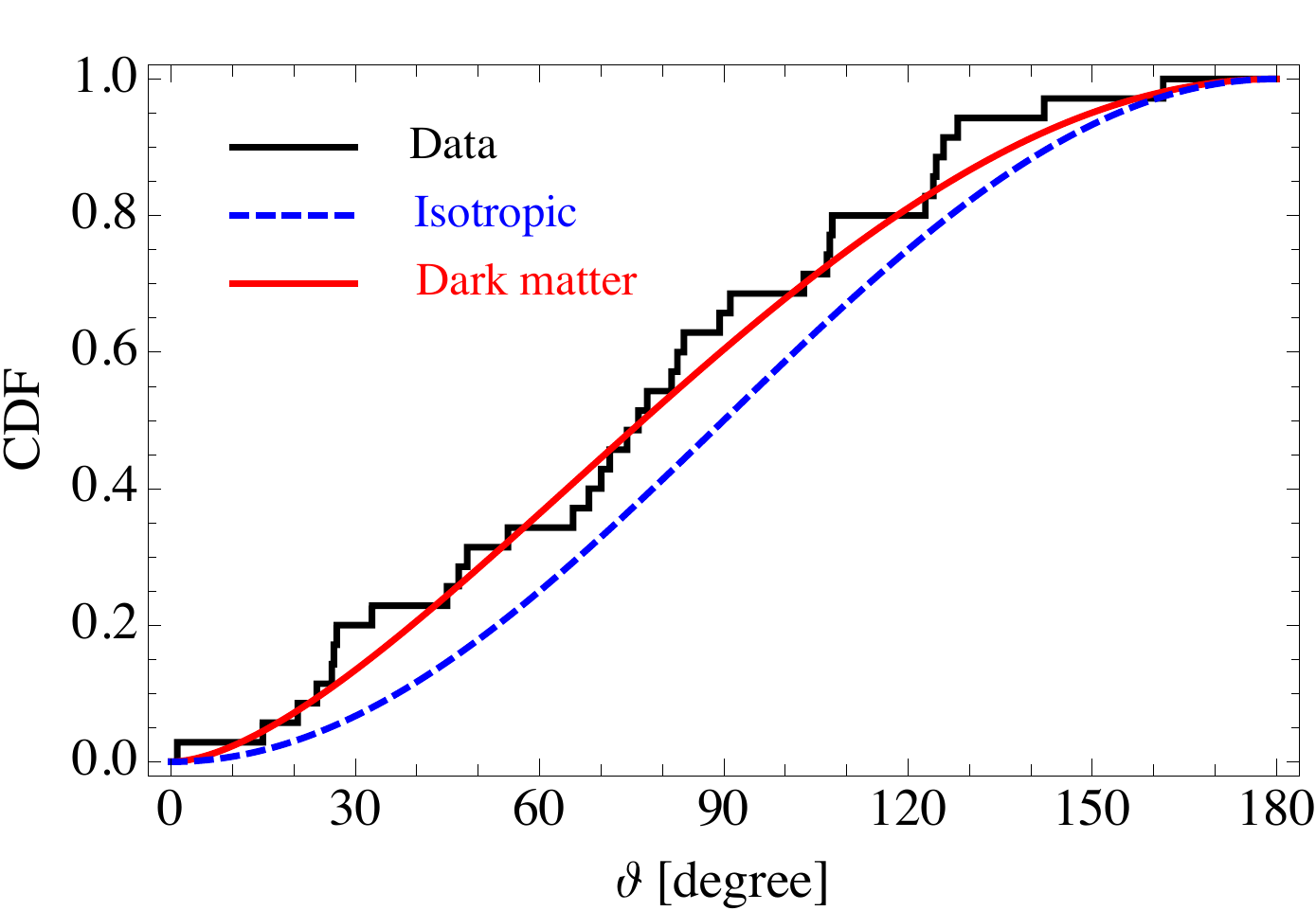}
\caption{\label{fig:cdf} Comparison of DM (red solid) and isotropic (blue dashed) CDFs with the EDF of IceCube data (black solid).}
\end{figure}

\begin{figure}[t!]
\centering
\subfloat[]{
\includegraphics[width=0.5\textwidth]{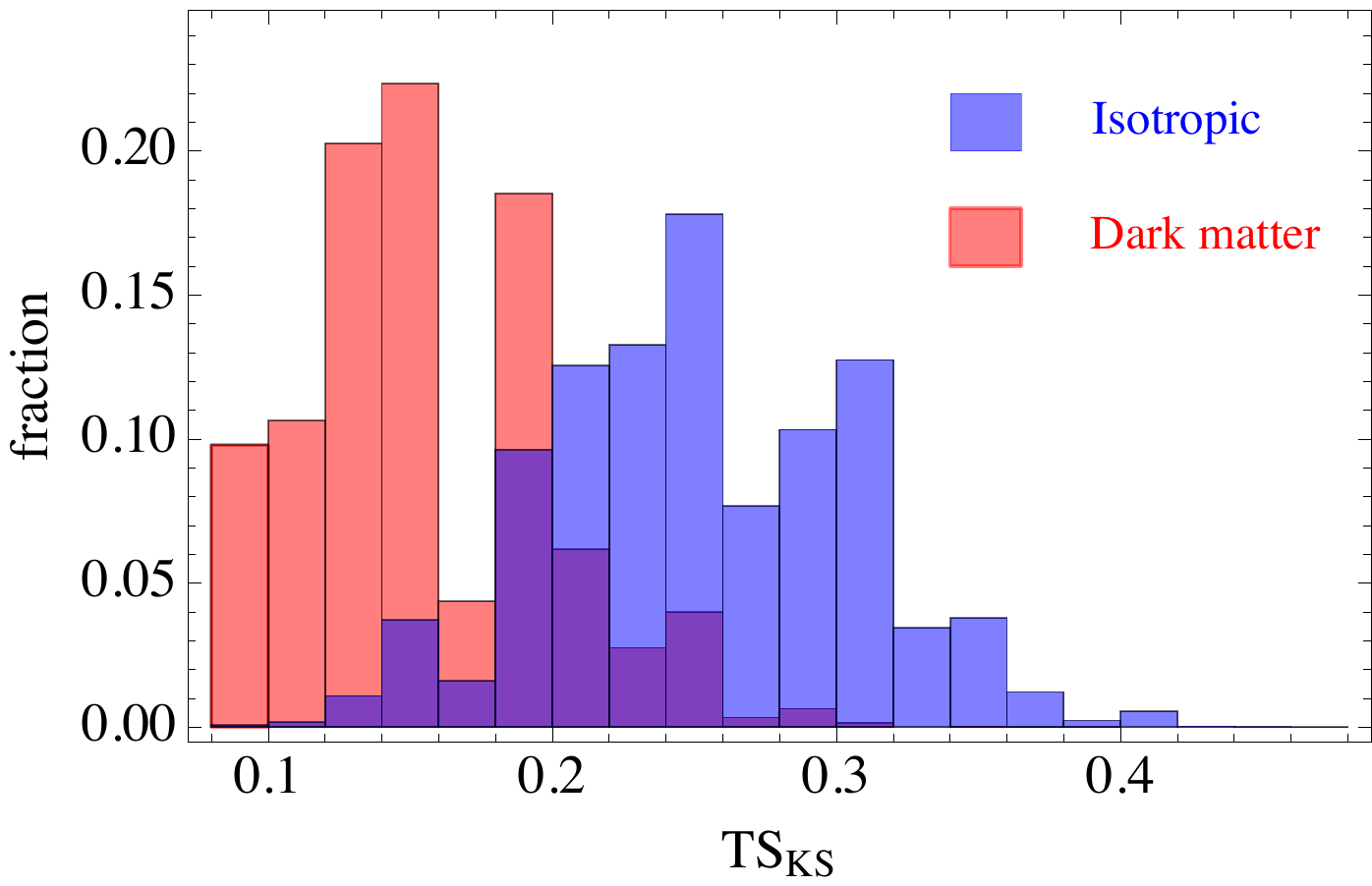}
\label{fig:KS-dis}
}
\subfloat[]{
\includegraphics[width=0.5\textwidth]{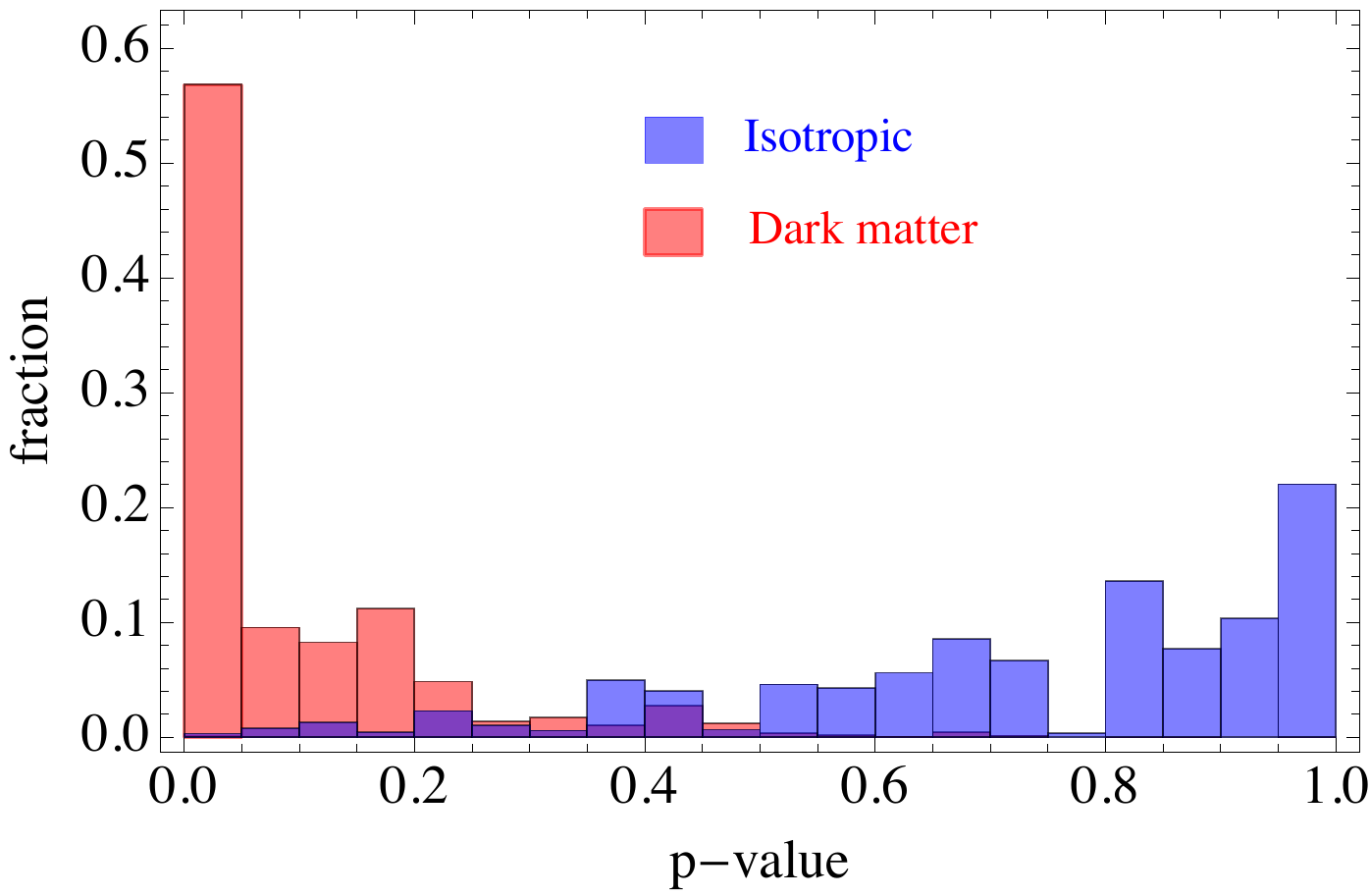}
\label{fig:KS-pvalue}
}
\caption{\label{fig:ks} Panel (a): histogram of the fraction of realizations out of $26 \choose 15$ having ${\rm TS}_{\rm KS}$ value within the shown bins. Panel (b): histogram of the fraction of realizations out of $26 \choose 15$ having $p$-values in the shown bins. In both panels the red (blue) color corresponds to comparison of EDF of data with DM (isotropic) CDF.}
\end{figure}

For illustration, Figure~\ref{fig:cdf} shows the CDF for DM (red solid) and isotropic (blue dashed) distributions, and EDF for {\it all the} data, i.e. including the background events. Graphically data show a preference for DM distribution; however, as we discussed in section~\ref{likeli}, the contribution of background events to the EDF should be taken into account. The statistical estimator used for the KS test consists in the maximal distance between the EDF and the theoretical CDF of tested distribution. For instance, for the case of DM the test statistics is defined as
 \begin{equation}
{\rm TS}_{\rm KS} = \max_{1\leq i\leq N} \left\{ {\rm CDF}^{\rm DM} (\vartheta_i) - \frac{i-1}{N} , \frac{i}{N} - {\rm CDF}^{\rm DM} (\vartheta_i) \right\}~.
\end{equation} 
An analogous definition holds for the isotropic case by replacing ${\rm CDF}^{\rm DM}\to{\rm CDF}^{\rm iso}$. To account for the fraction of background events, we follow the same procedure as for the likelihood test: we assume that the signal is made by $N=20$ events, and that all the 9 events with energy $> 150$ TeV are signal events, so that 15 out of 26 events with lower energies are background events. Then, we calculate the estimator TS$_{\rm KS}$ for all the possible ways of choosing 15 events out of 26 events. For the time being, we will use only the best-fit angular positions of the data. We shall comment on this approximation in section~\ref{otherissues}. Figure~\ref{fig:KS-dis} shows the distribution of ${\rm TS}_{\rm KS}$ values for DM and isotropic distributions, for all the possible ways of choosing background events. As can bee seen in Figure~\ref{fig:KS-dis}, ${\rm TS}_{\rm KS}$ have statistically smaller values for the DM case with respect to the isotropic one, which means that again the data  prefer to some extent the former distribution.

To calculate the $p$-value, we generate $\sim10^5$ sets of events (each set including 20 events) according to the isotropic distribution and for each set we calculate the ${\rm TS}_{\rm KS}$ for both DM and isotropic distributions. As in the likelihood test, for each realization of background choosing (that is ${26 \choose 15}$ ways) the $p$-value is calculated by comparing the ${\rm TS}_{\rm KS}$ of that realization with the ${\rm TS}_{\rm KS}$ distribution of generated events, that is the fraction of generated events having smaller ${\rm TS}_{\rm KS}$. Figure~\ref{fig:KS-pvalue} shows the distribution of $p$-values for all the realizations of ${26 \choose 15}$ ways. As can be seen, the two distributions are separated, with the DM one having statistically smaller values (hence being closer to the data). On the average, 10\% of generated isotropic sample have smaller ${\rm TS}_{\rm KS}$ than the values obtained for data vs. DM distribution, i.e. nine times out of ten  isotropically distributed events lead to a more distant cumulative distribution from the data. This can be interpreted as a mild (less than ``2 $\sigma$'') preference for the DM distribution, qualitatively similar to the likelihood test results in section~\ref{likeli} (for comparison, 73\% of generated isotropic sample have smaller ${\rm TS}_{\rm KS}$ than the values obtained for data vs. isotropic distribution.) The fact that the $p$-value is now higher than in the likelihood test is also reasonable, since the likelihood test uses more information and should thus be more powerful.

\subsection{Anderson-Darling test}\label{ADtest}

\begin{figure}[t!]
\centering
\subfloat[]{
\includegraphics[width=0.5\textwidth]{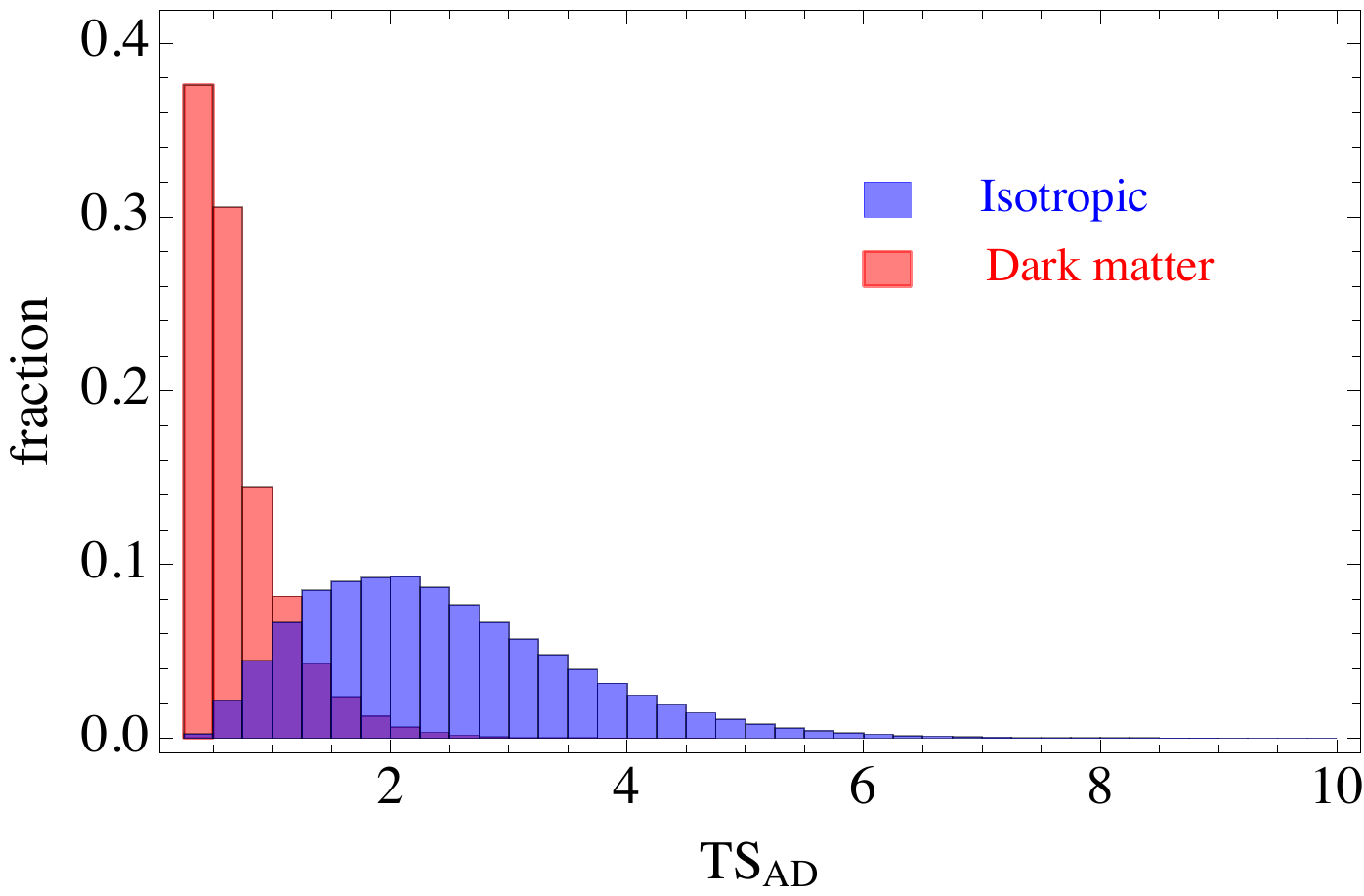}
\label{fig:AD-dis}
}
\subfloat[]{
\includegraphics[width=0.5\textwidth]{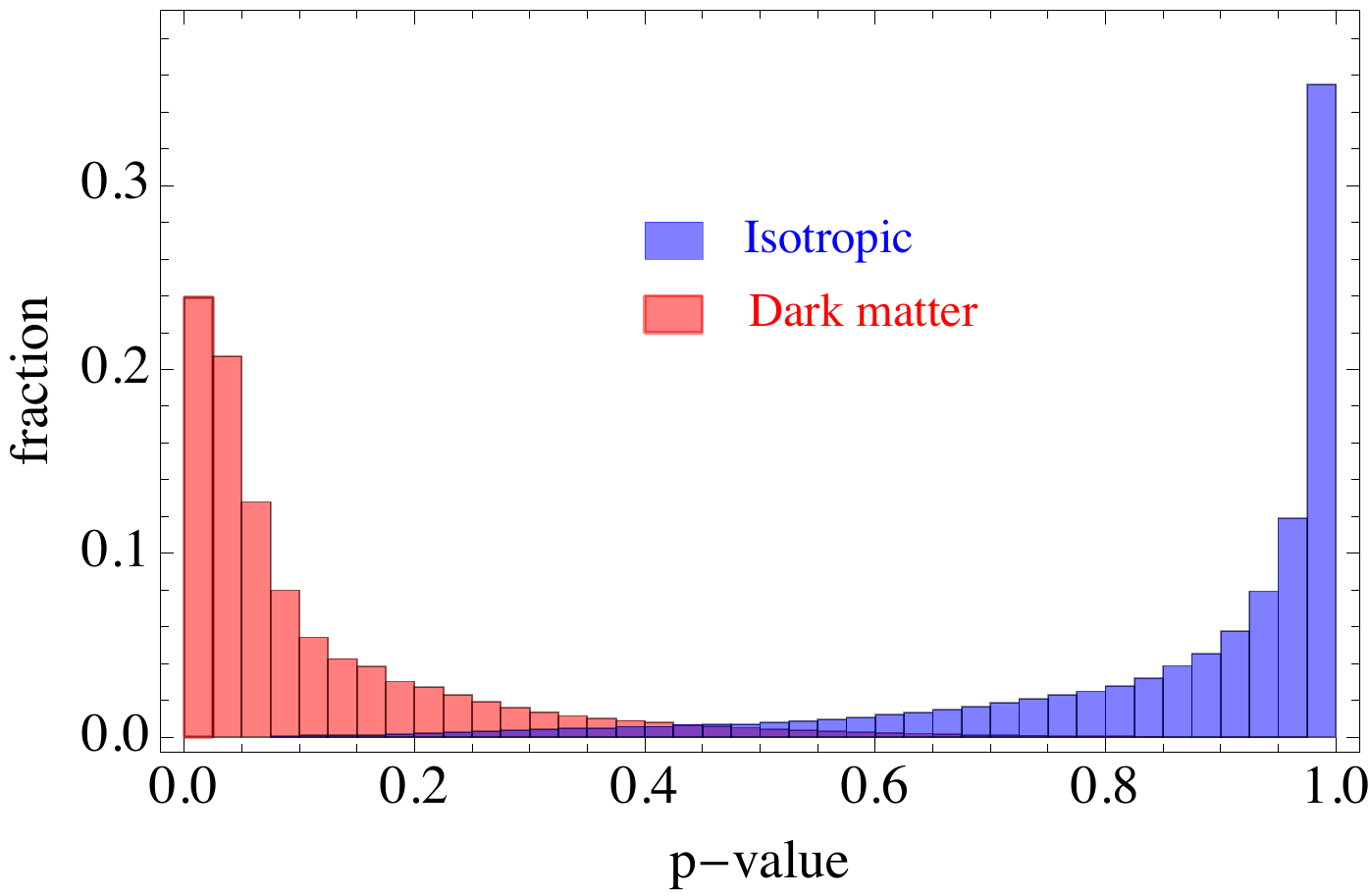}
\label{fig:AD-pvalue}
}
\caption{\label{fig:ad} The same as Figure~\ref{fig:ks}, but for AD test.}
\end{figure}

One of the shortcomings of the KS estimator used in the previous section is that generally it is not particularly sensitive to deviations of the EDF from the CDF at the endpoints (that is $\vartheta\simeq0$ and $\pi$), while the DM distribution differs from the isotropic one notably at $\vartheta\simeq0$. A powerful test in such cases is the Anderson-Darling (AD) test, which in its basic form used here is non-parametric. The AD test statistics estimator is defined as
\begin{equation}
{\rm TS}_{\rm AD} = -N -\frac{1}{N} \sum_{i=1}^N (2i-1) \left[ \ln \left({\rm CDF}^{\rm DM}(\vartheta_i)\right) + \ln \left( 1- {\rm CDF}^{\rm DM}(\vartheta_{N+1-i}) \right) \right]~,
\end{equation}
and similarly for the isotropic distribution. Here again $N=20$. Like the KS test, we calculate the ${\rm TS}_{\rm AD}$ for all the possible ways of dropping the 15 background events from the 35 events. Figure~\ref{fig:AD-dis} shows the distribution of ${\rm TS}_{\rm AD}$ for all the possible realization, for both DM and isotropic distributions, which the statistically larger ${\rm TS}_{\rm AD}$ for isotropic distribution indicates the preference of data for DM distribution.

Again, like the KS test, to calculate the $p$-value, we generate $\sim10^5$ sets of events (each including 20 events) according to the isotropic distribution. With the same method described in section~\ref{KStest} we calculate the distribution of $p$-values for AD test, which is shown in Figure~\ref{fig:AD-pvalue}. The separation of the DM and isotropic distributions under the AD test is visually stronger, with clearly smaller values of the estimator for the DM case with respect to isotropic one. This confirms the results of sections~\ref{likeli} and \ref{KStest}. Proceeding as in the previous case, we found results which are very similar to the KS case, namely that 11\% of generated isotropic sample have smaller ${\rm TS}_{\rm AD}$ than the values obtained for data vs. DM distribution (for comparison, 86\% of generated isotropic sample have smaller ${\rm TS}_{\rm AD}$ than the values obtained for data vs. isotropic distribution.)

\subsection{Other issues about the directional analysis}\label{otherissues}

In the tests performed previously, some simplifying approximations have been used. In this section we comment on a couple of these approximations and their effect on our results. Also, we perform a \textit{forecast} analysis which basically addresses the question of how many signal events are required to distinguish between the DM and isotropic distributions at a certain confidence level. This is a useful theoretical ``benchmark'' information to keep in mind, independent of experimental details which might have been mis-modeled in our treatment.  

\begin{itemize}
\item \underline{Angular Resolution}: in both the KS and AD tests, we neglected the resolution of the detector in reconstructing the direction of incoming neutrinos and we fixed the incoming direction of events to their best-fit values reported by IceCube. However, for cascade events there is a large uncertainty in the incoming direction which cannot be neglected. In principle, one can think of generalizing the previous method by increasing the number of Monte Carlo realizations further, replacing each generated data set with a distribution, with each data point replaced by a sampling over its gaussian error. Unfortunately, a quick estimate of the runs needed to correctly sample the distribution function is sufficient to convince oneself that it is numerically unfeasible.

On the other hand,  the simpler question if the ``preference'' for a DM-like distribution can be washed away by accounting for angular resolution can be addressed with a single and conservative academic case: we repeated the analysis by shifting all the incoming directions of cascade events $10^\circ$ up (that is $10^\circ$ farther from the GC). We  have performed the KS and AD tests on this \textit{shifted} data set, and the obtained level of discrimination is approximately similar to one reported in sections~\ref{KStest} and \ref{ADtest}. In more detail, the average $p$-value of KS and AD test for ``data vs. DM'' are close to the previous ones, while for ``data vs. isotropy'' average $p$-value decreases to $\sim40\%$; which means the shifted data are more compatible with the isotropic distribution (as we expected). However, the preference for the DM distribution persists.

\item \underline{Background Rejection}: in previous analyses, we simplified the ``signal/background'' separation by simply assuming that all the events with $E_\nu> 150$~TeV are signal, and being agnostic on the remaining ones. However, one can think of more clever separation criteria, like the method developed (and used in IceCube data analysis) in~\cite{Schonert:2008is,Gaisser:2014bja}, which combines a dominant veto criterion based on the declination of the events with some energy cut. If we use this method, one ends up inferring that 13 out of the 35 events are signal events\footnote{With the numbering according to SUPPL. TABLE I in~\cite{Aartsen:2014gkd}, these events are: 4, 12, 13, 14, 15, 17, 19, 20, 22, 26, 30, 33, 35.} and the 15 background events should then be selected among the 22 remaining events. Performing the statistical tests by this events selection, still one finds similar values for $p$-values as reported in sections~\ref{KStest} and \ref{ADtest}. This is not surprising, since a priori we expect that the high energy events, which are among the signal events in both methods of background rejection, play the crucial role in both analyses.

\item\underline{DM halo profile}: 
The DM signal depends on its distribution in the Galaxy, which has some uncertainties. However, altering it within reasonable ranges  only leads to minor changes to our angular template. This is due to the relatively mild angular dependence of the DM  signal in case of decay (as opposed to annihilation) together with fact that we perform a wide sky analysis (e.g. we do not concentrate on the few degrees around the Galactic Center).  For instance, by using the three templates of profiles reported in Ref.~\cite{Cirelli:2009dv}, the fraction of Galactic DM signal from the inner 30$~^\circ$ varies between $19\%$ (for a isothermal with a 3.2 kpc core) to $21\%$ (for the so-called Einasto profile), with our reference NFW profile in between, at about $20\%$. This is to be compared with less than $7\%$ of the total signal expected for an isotropic profile. These $5\%$ uncertainties are not worrisome given the present level of accuracy,  certainly subdominant with respects to some of the other effects discussed below. 

\item \underline{Forecast}: There are obviously further approximations not addressed in the above treatments. For instance, we did not allow for a variable level of background, neither for the instrumental dependence of the effective area on the angular direction. In fact, there are reasons to believe that accounting for the latter effect would not alter much our previous conclusions, since at least for events with declination lower than 20$^\circ$-30$^\circ$, the IceCube detector would actually see an incoming isotropic flux as almost isotropic (see for instance the solid and dashed gray lines in Fig.~3 of Ref.~\cite{Aartsen:2014gkd}). Only a handful of data are characterized by higher declinations, where the Earth acts as a shield for higher energy events. 

\begin{figure}[t!]
\centering
\includegraphics[width=0.5\textwidth]{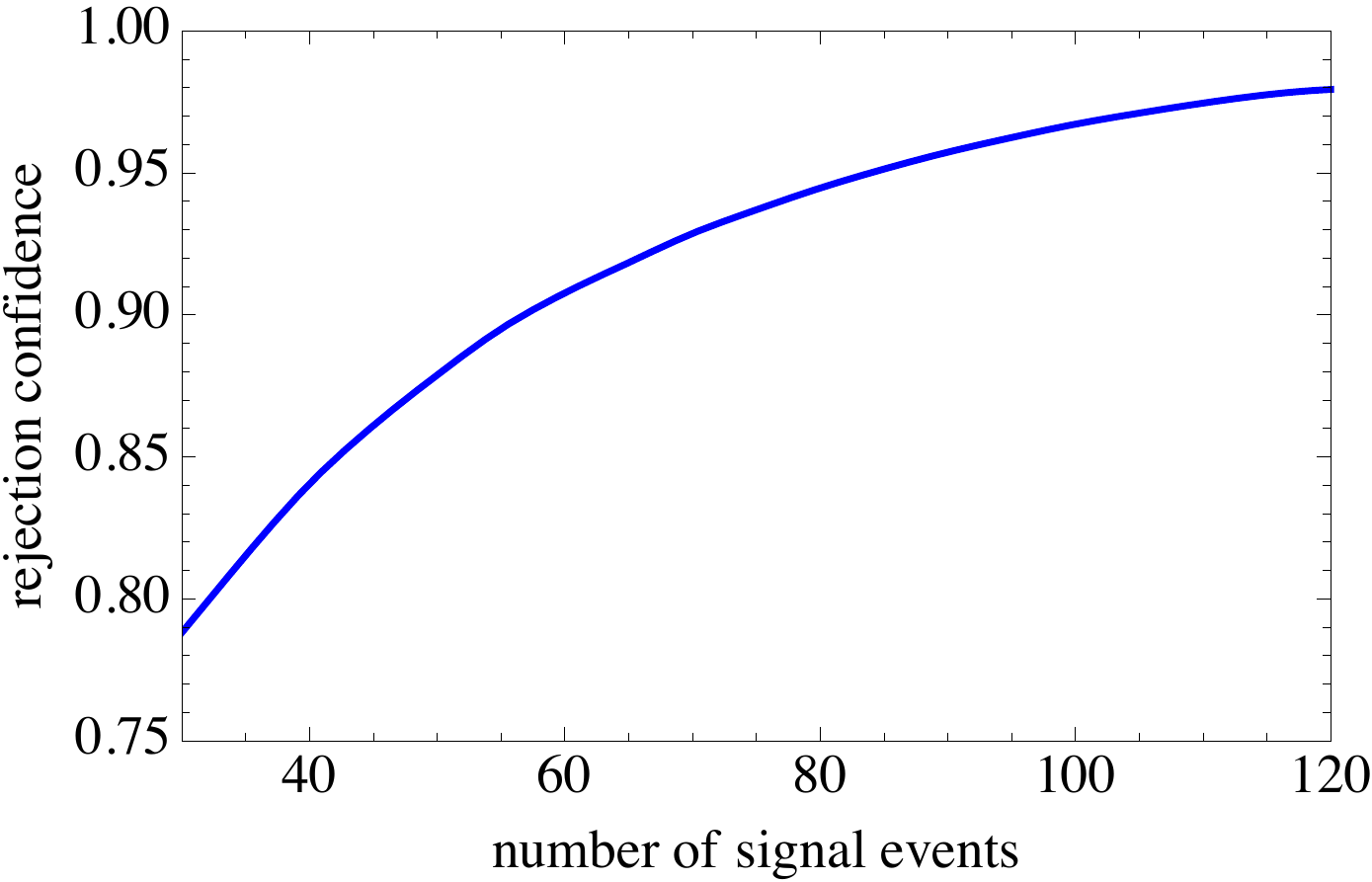}
\caption{\label{fig:prospect} Rejection confidence level of a DM-like distribution given isotropically distributed data, as function of the number of events.}
\end{figure}

A further sanity check consists in asking oneself if the results we obtained are reasonable, based only on statistical expectations rather than detector-dependent considerations and the actual data. For instance: on a purely theoretical ground, how many events, if isotropically distributed, are needed in order to reject a DM-like anisotropy pattern at a certain confidence level? This is also useful to gauge the diagnostic potential that can be reasonably expected by the statistics collected by IceCube over its whole lifetime of about $20$ years.

For the specific case of the KS test, the critical value of ${\rm TS}_{\rm KS}$ for a confidence level $1-\alpha$ (provided that  $n\gtrsim 35$) is given analytically by $\sqrt{\frac{2}{n}\ln(\frac{2}{\alpha})}$. We generate $n$ events according to the isotropic distribution and then we calculate at which confidence level it is possible to reject the DM distribution by calculating the $p$-value by comparing to $\sim10^5$ generated data set of events. Figure~\ref{fig:prospect} shows the rejection confidence level as function of the number of events. A current diagnostic potential around 80\% C.L. appears reasonable, comforting our previous analyses. Also, it suggests that with statistics of about $120$ events (which is reasonably within reach of the lifetime of IceCube detector) it should be possible to reject the DM distribution at about 99\% C.L. 

\end{itemize}

\section{Dark Matter decay, energy spectra and possible production mechanisms}
\label{sec:models}

Compared with the angular distribution of the events, most of the features of the neutrino spectrum from DM decay are significantly more model-dependent. As we argued in our previous publication~\cite{Esmaili:2013gha}, the most generic feature is that the spectrum presents a high-energy cutoff at some fraction of the DM mass, typically $m_{\rm DM}/2$ if two body decays including one neutrino are present. Differently from most astrophysical models which predict relatively featureless spectra over several decades of energy, spectra associated to DM decay are expected to depart somewhat from a (scale invariant) power-law. The most common spectral feature is some ``spectral dip'' between the end-point and the low-energy part of the spectrum, where the atmospheric (and possibly other astrophysical) backgrounds take over. This feature is typically expected in all (relatively common) circumstances where the spectrum is characterized by comparable level of ``soft'' neutrino channels (as coming from quark cascades) and ``hard'' neutrino channels (notably associated to leptonic final states). This was shown parametrically for different final states in~\cite{Esmaili:2013gha}. Here we take a more theoretically justified approach, but shall limit ourselves to some benchmarks to illustrate the main points. 

In a large class of models, named of the ``portal type'', a link between the dark sector and the standard model can be established via a Lagrangian of the form
\begin{equation}
{\cal L}_{\rm portal}=\frac{{\cal O}_{\rm SM}{\cal O}_{\rm DM}}{\Lambda^{d-4}}\,,
\end{equation}
where ${\cal O}_{\rm SM}$ and ${\cal O}_{\rm DM}$ are gauge-invariant operators composed solely of the standard model and dark sector fields, respectively, $d$ is the dimension of the (in general, effective field theory) portal operator, and $\Lambda$ is an energy scale. The so-called ``neutrino portal''~\cite{Falkowski:2009yz} (see also~\cite{Anisimov:2008gg} for earlier related ideas) consists of replacing $\mathcal{O}_{\rm SM}$ with the singlet operator $(H\,L)$ constructed from the lepton and Higgs SU(2) doublets. In the standard model this is the lowest dimensional (dimension 5/2) singlet operator that can be used to build a model where the DM candidate decays into neutrinos.

\begin{figure}[t!]
\centering
\includegraphics[width=0.6\textwidth]{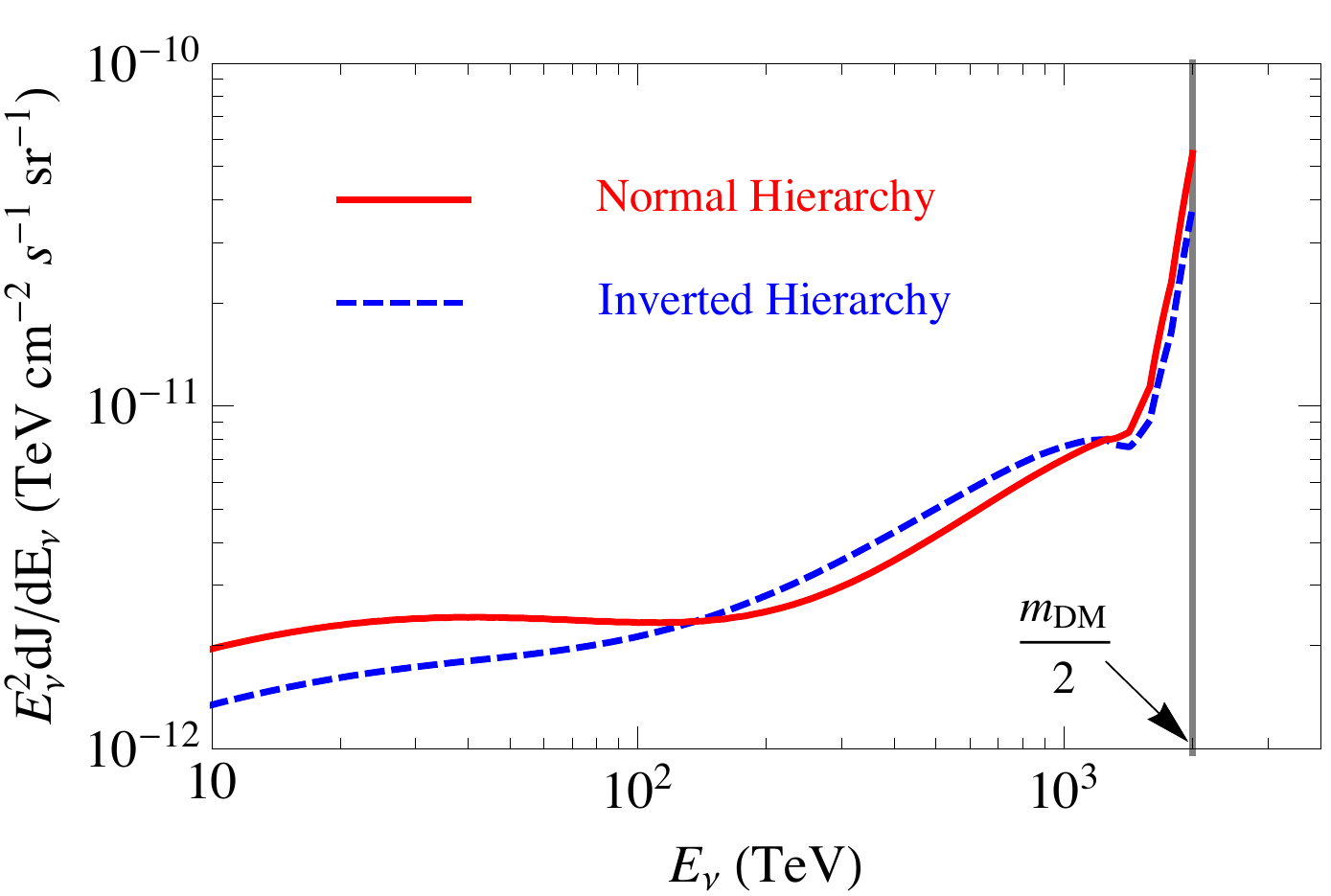}
\caption{\label{fig:spec}The energy spectrum of $(\nu_e+\nu_\mu+\nu_\tau)/3$ from decaying DM of the model proposed in~\cite{Higaki:2014dwa}, for NH and IH cases. For the mass of DM we assumed $m_{\rm DM}=4$~PeV, and for lifetime: $7.3\times10^{27}$~s for NH and $1.1\times10^{28}$~s for IH.}
\end{figure}

Concerning ${\cal O}_{\rm DM}$, the simplest possibility is obviously to replace it with a spin-1/2 gauge singlet neutrino $N$ identified with the DM candidate. Basically this corresponds to DM being a heavy ``sterile neutrino'', in which case ${\cal L}_{\rm portal}$ has dimension $d=4$ and the corresponding theory is renormalizable. This possibility has been explored in~\cite{Higaki:2014dwa} in connection with the interpretation of the IceCube data (see also~\cite{Anisimov:2008gg} for an earlier model appearing well before the IceCube detection, sharing some of these features). Since this is the simplest case, let us summarize the main results: The lifetime of $N$ (which is the DM candidate) is roughly given by~\cite{Higaki:2014dwa} 
\begin{equation}
\tau_{\rm DM}\sim 8\times 10^{28}\;\, {\rm s} \;\left(\frac{M_N}{1\;{ \rm PeV}}\right)^{-1} \frac{10^{-29}}{\left|y_{N}\right|^2}~,
\end{equation} 
where $M_N$ is the mass of $N$ and $y_N$ is the dimensionless coefficient of the operator $N\, L\, H$. At tree level, $N$ decays to $\ell^\pm W^\mp$, $\nu_\ell Z$ and $\nu_\ell h$ channels with the relative branching ratios $2:1:1$. In the specific model of Ref.~\cite{Higaki:2014dwa} even the flavor structure of decay channels ({\it i.e.}, the branching ratios to each flavor) are determined. To be definite, let us separately consider the two cases, normal mass hierarchy (NH) and inverted mass hierarchy (IH), as reported in~\cite{Higaki:2014dwa}, which lead to the following branching ratios: 

\begin{itemize}
\item \textit{Normal Hierarchy} (NH): ${\rm Br}(\ell^\pm W^\mp) = 2\,{\rm Br}(\overset{(-)}{\nu_\ell} Z) = 2\,{\rm Br}(\overset{(-)}{\nu_\ell} h) = \left|U_{\ell1}\right|^2~,$
\item \textit{Inverted Hierarchy} (IH): ${\rm Br}(\ell^\pm W^\mp) = 2\,{\rm Br}(\overset{(-)}{\nu_\ell} Z) = 2\,{\rm Br}(\overset{(-)}{\nu_\ell} h) = \left|U_{\ell3}\right|^2~,$ 
\end{itemize}
where $ \ell=e,\mu,\tau$ and $U_{\ell i}$ are the elements of the neutrino mixing matrix (we assume CP conservation in neutrino sector). For the mass of DM particle we have the freedom to choose a value within {\it twice} the energy width of last (observed) bin of IceCube data. We choose $m_{\rm DM}\equiv m_N=4$~PeV. Figure~\ref{fig:spec} shows the energy spectrum of $(\nu_e+\nu_\mu+\nu_\tau)/3$ (relying on~\cite{Cirelli:2010xx} computed as in~\cite{Esmaili:2013gha}) for this mass for both NH and IH cases.  

In this model, a variety of production mechanisms are possible depending on the details of its completion. For example, by adding a mass term $M\,N\,N$ and an interaction term with a scalar $\Phi$, $g\,\Phi\,N\,N$, we can consider the possibility that $N$ is populated via the decay of $\Phi$ in the case that $\Phi$ is heavier than $N$. This possibility has been studied in~\cite{Higaki:2014dwa} by taking $\Phi$ to be the ``Higgs'' of the broken gauged $B-L$ symmetry. In this scenario, $\Phi$ can also be invoked for both inflation and leptogenesis, and the mass of $\Phi$, $m_{\Phi}$ is acquired via the vev of $\Phi$. We note however that this is by no means the unique choice. For instance, for the case $m_\Phi \ll m_N$, the sterile neutrino can attain the right abundance via the so-called ``freeze-in scenario'' for coupling $g\simeq 10^{-6}$; this would represent a variation of the model described in~\cite{Klasen:2013ypa}.

Another possible choice for the ``Dark Sector'' operator would be to replace ${\cal O}_{\rm DM}$ with a gauge-invariant combination (mimicking the SM operator $(H\,L)$) built out of a singlet scalar $\phi$ and a singlet fermion $\chi$; {\it i.e.}, ${\cal O}_{\rm DM}\to \chi\,\phi$, which leads to the unique dimension 5 operator of this portal type. The spectra would be different in this case due to the presence of a ``dark'' daughter particle in the final state. This operator is typically adopted in ``Asymmetric Dark Matter'' models, see for instance the review~\cite{Zurek:2013wia}, or~\cite{Falkowski:2011xh}  for a specific example. These scenarios imply a link between the abundance of DM and the generation of baryon asymmetry. Although it is possible certainly to utilize these models to allow for a PeV-scale DM, we do not indulge in further details here since phenomenologically one can reproduce signals similar to the one mediated by the dimension four operator, apart from the kinematical differences.

\begin{figure}[t!]
\centering
\subfloat[NH]{
\includegraphics[width=0.5\textwidth]{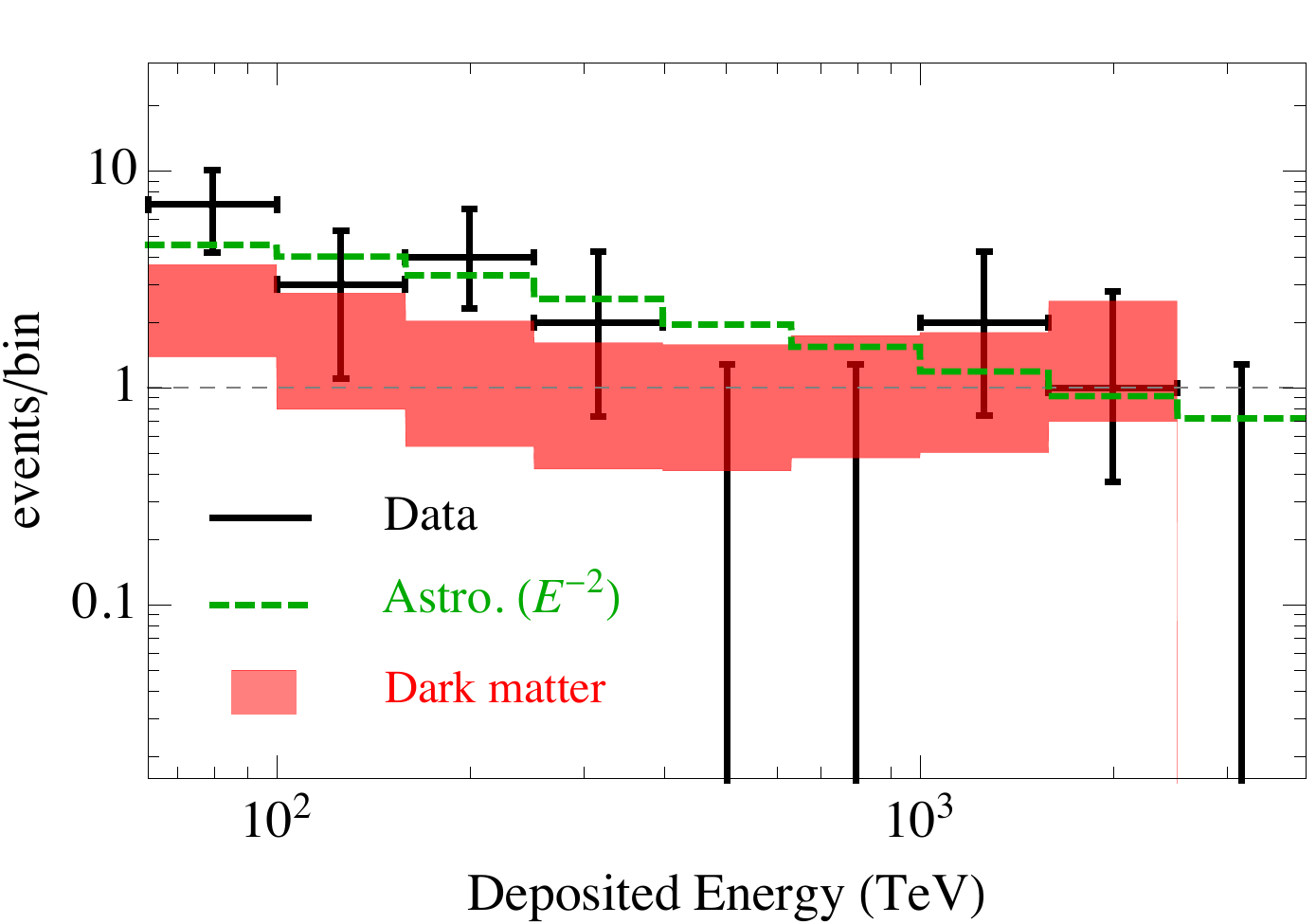}
\label{fig:eventsNHIH}
}
\subfloat[IH]{
\includegraphics[width=0.5\textwidth]{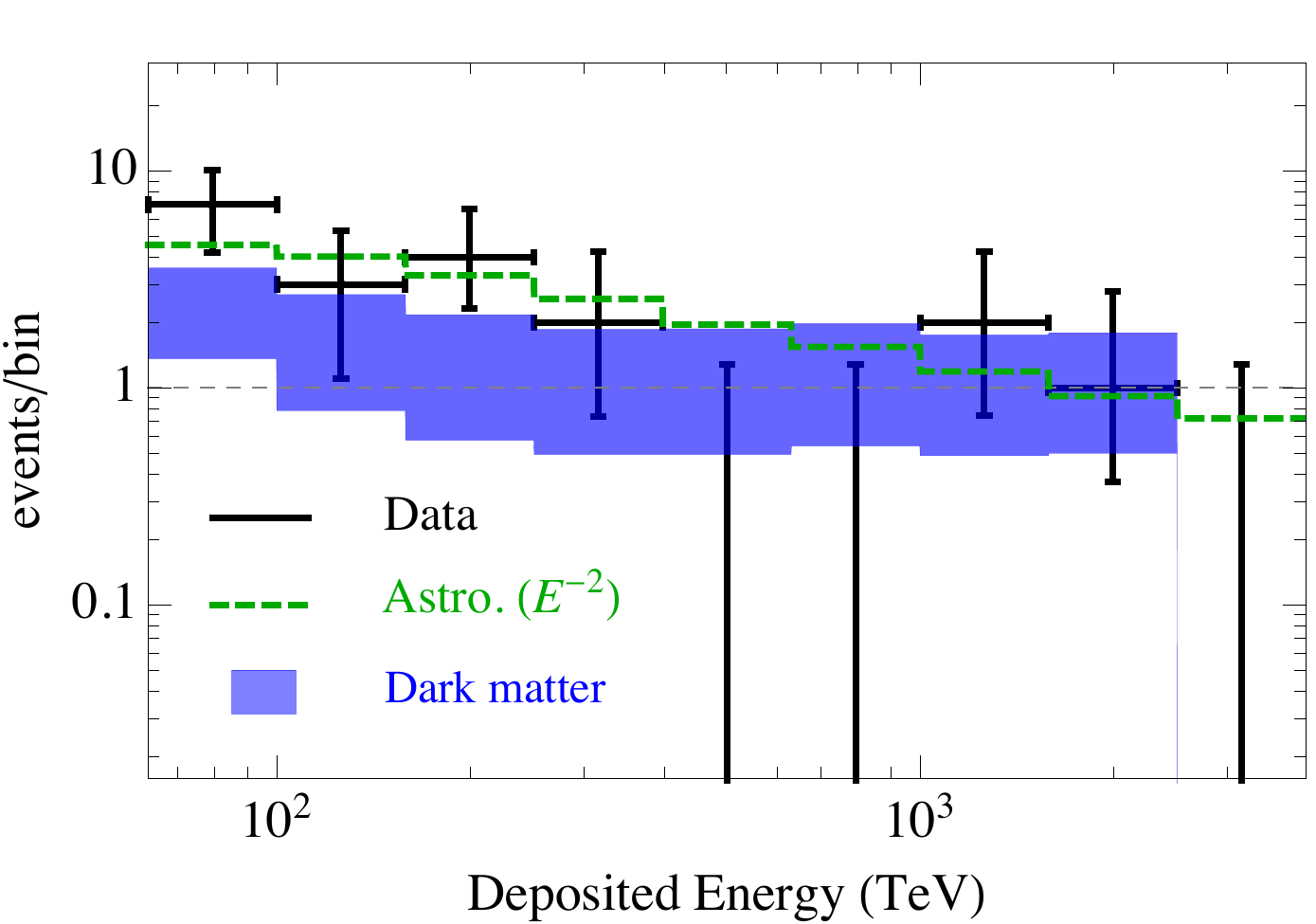}
\label{fig:events-example}
}
\caption{\label{fig:events} The shaded regions show the energy distribution of events for NH (left panel) and IH (right panel). The width of the shaded regions corresponds to variation of $\tau_{\rm DM}$ within $1\sigma$ range around the best-fit point obtained from the fit to IceCube data ($\tau_{\rm DM}=7.3\times10^{27}$~s for NH and $\tau_{\rm DM}=1.1\times10^{28}$~s for IH). For both NH and IH we assume $m_{\rm DM}=4$~PeV. The green dashed line shows the expected events from astrophysical $E_\nu^{-2}$ flux. The black points show the IceCube data. }
\end{figure}

At dimension 6 and/or higher, a larger number of operators containing other ``portals'' in the standard model become possible. A list of those carrying $B-L$ quantum number is reported for instance in~\cite{Feldstein:2010xe}. The discussion becomes soon very technical and model-dependent. Qualitatively, however, the higher-dimensional operator models share the possibility of sizable branching ratios into hadronic final states, associated to softer neutrino spectra. Thus, while any model that fits the spectrum should be relatively similar in the hard channel (associated to ``leptonic'' operators), the lower-energy part of the spectrum is more model-dependent. In~\cite{Esmaili:2013gha}, we already argued that with both soft and hard channels available, a fit of the data can be easily achieved. It is worth showing however that already with the most constrained model, corresponding to the dimension-4 operator $N\,H\,L$, an acceptable fit can be obtained: Figure~\ref{fig:events} shows the energy distribution of events for a 4 PeV DM mass, including both the DM signal and the atmospheric residual background (only relevant at low energies), for the previously mentioned NH model (left panel) and IH model (right panel). The width of the shaded regions corresponds to variation of $\tau_{\rm DM}$ within $1\sigma$ range around the best-fit point obtained from the fit to IceCube data ($\tau_{\rm DM}=7.3\times10^{27}$~s for NH and $\tau_{\rm DM}=1.1\times10^{28}$~s for IH). The dashed green line shows the expected events for an $E_\nu^{-2}$ astrophysical spectrum. Qualitatively, this DM model (the simplest one!) is in a better  agreement with the data than the astrophysical one above about 300 TeV, and in slightly worse agreement at lower energies. However, in our opinion the model matches the data  very satisfactorily, given the fact that it is very reasonable to expect either some extra signal at low energy (coming from sub-leading soft channels of DM decay, as for instance mediated by some dimension-6 operators) or extra background from either prompt atmospheric neutrinos or other astrophysical neutrinos with a relatively steep spectrum.

\section{Gamma-ray bounds}\label{sec:gamma}

Any high-energy neutrino signal is associated with gamma-ray signals. The DM decay is not an exception and here we discuss some of these potential signatures. At energies $E_\gamma> \,$TeV, the Universe is opaque to gamma-rays. However, cascades develop: gamma-rays produce $e^\pm$ pairs through the interaction with the interstellar radiation field and, at PeV energies, with the CMB. In turn, these $e^\pm$ will produce further gamma-rays via inverse Compton onto CMB photons. The process continues until gamma-ray energies fall below a threshold of ${\cal O}(100)\,$GeV. So, the extragalactic component of the DM electromagnetic signal ({\it i.e.}, gamma and $e^{\pm}$ fluxes) will certainly end up in cascades. The inverse Compton scattering photons induced by  $e^{\pm}$ will populate a gamma-ray flux with energies roughly between $E_1\sim {\cal O}(1)$ GeV and $E_2 \sim{\cal O}(100)$ GeV, see for example Fig.~5 in~\cite{Semikoz:2003wv}. This flux is constrained by the Fermi data~\cite{Ackermann:2014usa} which have an {\it integrated} energy density (figure affected by a 20\%-30\% uncertainty)
\begin{equation}\label{eq:fermi}
\omega_\gamma\,=\,\frac{4\pi}{c}\int_{E_1}^{E_2}
E_\gamma\, \frac{{\rm d} \varphi_{\gamma}}{{\rm d}E_\gamma}\, \mathrm{d}E_\gamma\, \lesssim\, 4.4\times 10^{-7} \,\textrm{eV/cm}^{3}~\,.
\end{equation}

To check if the expected gamma-ray signal from decaying DM with properties discussed in section~\ref{sec:models} violates the bound in eq.~(\ref{eq:fermi}), we calculate the total electromagnetic energy budget; {\it i.e.}, the total energy density deposited in prompt gamma and $e^{\pm}$ in the decay of DM. A quick computation shows that for the NH case in the previous section, one obtains 
\begin{equation}
\frac{4\pi}{c}\int \left[\left. E_\gamma\,\frac{{\rm d} \varphi_{\gamma}}{{\rm d}E_\gamma}\right)^{\rm extragal}+ E_e\, \left.\frac{{\rm d} \varphi_{e^\pm}}{{\rm d}E_e}\right)^{\rm extragal}\right]\, \mathrm{d}E\, \simeq\, (1.0 +1.5)\times 10^{-8} \;\textrm{eV/cm}^{3}~, 
\end{equation}
and for IH case, the energy density is smaller by a factor of two. The above energy density is smaller than the one measured by the Fermi-LAT  by almost a factor of $\sim20$. Even including the residual isotropic contribution from DM halo (i.e. the anti-GC flux), the energy density still falls below the Fermi bound by almost one order of magnitude
\begin{equation}\label{eq:gamma-gal}
\frac{4\pi}{c}\int
\sum_{{\rm i}={\rm gal,extragal}} \left[E_\gamma\, \left.\frac{{\rm d} \varphi_{\gamma}}{{\rm d}E_\gamma}\right)^{\rm i}+ E_e\, \left.\frac{{\rm d} \varphi_{e^\pm}}{{\rm d}E_e}\right)^{\rm i}\right]\, \mathrm{d}E\, \simeq\, 5.2\times 10^{-8} \;\textrm{eV/cm}^{3} \, . 
\end{equation}
The above figures imply that it is not possible to probe PeV-scale DM decay with Fermi data on the isotropic residual flux, yet. However, there is some mounting evidence that at least a significant fraction of the diffuse gamma-ray flux at high-energy may be due to unresolved BL Lac objects, see for instance the recent~\cite{DiMauro:2013zfa} and references therein. If this can be solidly established, a corresponding improvement of the constraining power can be expected.
  
In fact, for the Galactic contribution that we considered in eq.~(\ref{eq:gamma-gal}) the situation is more complicated and, at the same time, more promising since additional diagnostics is available. The absorption length of gamma-rays at PeV energies is comparable to typical Galactic distances: so, one cannot assume either full absorption/cascade development or complete transparency. Also, both the CMB and the Galactic interstellar radiation field are important (see for instance~\cite{Moskalenko:2005ng}). Both the computations of the non-absorbed gamma-ray flux and of the IC signal from diffuse photons up-scattered by $e^\pm$ from DM decay constitute in general complicated 3D problems. However, for our goal here it is sufficient to show that even the direct bounds on the photon flux at $\sim$~PeV energies are not ruling out the scenario considered in this paper (after having shown above that the cascade bounds are not constraining either). To that purpose, we compare the photon flux due to the Galactic DM halo with the CASA-MIA bounds of Ref.~\cite{Chantell:1997gs}. Since the CASA-MIA bounds were derived under the isotropic approximation, we compare the bounds with the anti-GC contribution of the halo signal, which represents the isotropic residual of the DM signal (the additional signal is indeed peaked towards the Galactic Center). On the other hand, to be conservative, we shall assume  {\it unattenuated} flux for the DM photons. The CASA-MIA bounds state that: 
\begin{itemize}
\item The integrated flux of $\gamma$-rays above 330~TeV must be below $1.0\times 10^{-13}\,{\rm cm}^{-2}\,{\rm s}^{-1}\,{\rm sr}^{-1}$.
\item The integrated flux of $\gamma$-rays above 775 TeV must be below $2.6\times 10^{-14}\,{\rm cm}^{-2}\,{\rm s}^{-1}\,{\rm sr}^{-1}$.
\end{itemize}
The integrated fluxes of gamma-rays, in the DM decay scenario previously discussed, above 330~TeV and 775 TeV are respectively $1.2\times 10^{-14}\,{\rm cm}^{-2}\,{\rm s}^{-1}\,{\rm sr}^{-1}$ and $8.8\times 10^{-15}\,{\rm cm}^{-2}\,{\rm s}^{-1}\,{\rm sr}^{-1}$, which again are three to eight times below the CASA-MIA bounds. 

For the time being, the above results show that the assumed DM decay scenario is safely within the $\gamma$ limits. However, more constraining result would follow from a ``customized'' analysis exploiting the peculiar angular dependence of the DM signal, as done e.g.~for the Galactic Plane profile template in~\cite{Borione:1997fy}.

\section{DM lifetime constraints from astrophysical explanations of IceCube data}
\label{sec:lifetime}

Up to this point we discussed the possibility of explaining the IceCube data within the decaying DM scenario. However, let us assume that the observed events in IceCube originate from yet unspecified astrophysical sources. In this case, IceCube observation can be used to constrain DM properties. In this section we derive the corresponding limits on heavy DM decay lifetime. Additionally, the lower energy part of IceCube data can be used to probe the annihilation cross section of DM. We study the annihilating DM case in appendix~\ref{sec:ann}.

A lower limit on the DM lifetime $\tau_{\rm DM}$ can be obtained with a method similar to the one used in~\cite{Esmaili:2012us}, to which we address the reader for further details. The major differences with respect to that paper consist in the more refined spectra (which account for $W,Z-$strahlung corrections) and the dataset that we are using in this section. In our analysis we consider the 12 bins of energy from $10^{1.4}$~TeV to $10^{3.8}$~TeV of IceCube data (see Figure~2 of Ref.~\cite{Aartsen:2014gkd}). The number of observed events in the $i$-th bin, $N_{\rm data}^i$, is given in the fourth column of Table~\ref{tab:numbers}.

\begin{table}[h]
\caption{Summary of the binned information used in our analysis for deriving limit on DM lifetime, including: bin width, the expected number of astrophysical events for $E_\nu^{-2}$ and $E_\nu^{-2.3}$ fluxes, $N_{\rm astro}$, the number of observed events, $N_{\rm data}$, and the upper limit on neutrino events at 90\% C.L., $N_{\rm limit}$, assuming $E_\nu^{-2}$ and $E_\nu^{-2.3}$ fluxes. The last column is the astro-model-independent values of $N_{\rm limit}$.}
\begin{center}
\begin{tabular}{|c|c|c|c|c|c|}
\hline
 bin \# & $\log_{10}(E_\nu/{\rm TeV})$ & $N_{\rm astro} (E_\nu^{-2} \div E_\nu^{-2.3})$ & $N_{\rm data}$  & $N_{\rm limit}$ ($E_{\nu}^{-2} \div E_{\nu}^{-2.3}$) & $N_{\rm limit}$ \\
 \hline
\#1 & $1.4-1.6$ & 9.46 $\div$ 10 &11 & 7.8 $\div$ 7.46 & 16.6 \\
\hline
\#2 & $1.6-1.8$ & 4.31 $\div$ 5.3 & 6 & 6.53 $\div$ 5.87 & 10.5 \\
\hline
\#3 & $1.8-2.0$ & 4.55 $\div$ 5.68 & 7 & 7.41 $\div$ 6.58 & 11.8 \\
\hline
\#4 & $2.0-2.2$ & 3.97 $\div$ 4.82 & 3 & 3.98 $\div$ 3.73 & 6.68 \\
\hline
\#5 & $2.2-2.4$ & 3.32 $\div$ 3.56 & 4 & 5.15 $\div$ 5.01 & 8.00 \\
\hline
\#6 & $2.4-2.6$ & 2.59 $\div$ 2.42 & 2 & 3.65 $\div$ 3.71 & 5.32 \\
\hline
\#7 & $2.6-2.8$ & 1.96 $\div$ 1.62 & 0 & 2.3 $\div$ 2.3 & 2.3 \\
\hline
\#8 & $2.8-3.0$ & 1.55 $\div$ 1.1 & 0 & 2.3 $\div$ 2.3 & 2.3 \\
\hline
\#9 & $3.0-3.2$ & 1.2 $\div$ 0.74 & 2 & 4.31 $\div$ 4.64 & 5.32 \\
\hline
\#10 & $3.2-3.4$ & 0.92 $\div$ 0.5 & 1 & 3.3 $\div$ 3.51 & 3.89 \\
\hline
\#11 & $3.4-3.6$ & 0.73 $\div$ 0.35 & 0 & 2.3 $\div$ 2.3 & 2.3 \\
\hline
\#12 & $3.6-3.8$ & 1.72 $\div$ 0.76 & 0 & 2.3 $\div$ 2.3 & 2.3 \\
\hline
\end{tabular}
\end{center}
\label{tab:numbers}
\end{table}

For the astrophysical neutrinos, we assume two unbroken power-law fluxes: $E_\nu^{-2}$ and $E_{\nu}^{-2.3}$ fluxes with the normalizations reported in IceCube paper~\cite{Aartsen:2014gkd}. The expected astrophysical events (including the best-fit atmospheric background) in each bin is given in the third column of Table~\ref{tab:numbers}. For each bin $i$ we can derive the $N_{\rm limit}^i$ such that the number of events from DM decay should be smaller than $N_{\rm limit}^i$. At $q\%$ confidence level, the $N_{\rm limit}^i$ can be read from the following equation:
\begin{equation}
\frac{q}{100} = \frac{\int_{0}^{N_{\rm limit}^i} L(N_{\rm data}^i,N)~{\rm d}N}{\int_{0}^{\infty} L(N_{\rm data}^i,N)~{\rm d}N}\,,
\end{equation} 
where
\begin{equation}\label{eq:L}
L(N_{\rm data}^i,N) = \frac{(N+N_{\rm astro}^i)^{N_{\rm data}^i}}{N_{\rm data}^i !} e^{-(N+N_{\rm astro}^i)}\,.
\end{equation}
The obtained $N_{\rm limit}^i$ for each bin, for both $E_\nu^{-2}$ and $E_{\nu}^{-2.3}$ astrophysical fluxes can be found in the fifth column of Table~\ref{tab:numbers}. Not surprisingly, the limits are comparable in the two cases of $E_\nu^{-2}$ and $E_{\nu}^{-2.3}$ fluxes, with differences typically at the 10\% level or smaller. This means that the limits are data-driven and robust with respect to the details of the astrophysical explanation involved. In particular, at higher masses, a similar result would be obtained if we had been even more agnostic on the astrophysical explanation and had simply derived, in each bin, the maximum theoretical flux consistent with the observed data. This astro-model-independent limit corresponds to replacing $L(N_{\rm data}^i,N)$ in eq.~(\ref{eq:L}) by  
\begin{equation}\label{eq:L-in}
L(N_{\rm data}^i,N) = \frac{(N)^{N_{\rm data}^i}}{N_{\rm data}^i !} e^{-N}~.
\end{equation}
The obtained $N_{\rm limit}^i$ for this case are shown in the last column of Table~\ref{tab:numbers}. As can be seen, for energies $\gtrsim400$~TeV the values of $N_{\rm limit}$ are almost the same as the case of $E_\nu^{-2}$ (or $E_\nu^{-2.3}$) astrophysical flux. Deviations grow at lower energies, since the exact amount of the signal accounted for by astrophysical sources becomes more important.

\begin{figure}[t!]
\centering
\subfloat[]{
\includegraphics[width=0.5\textwidth]{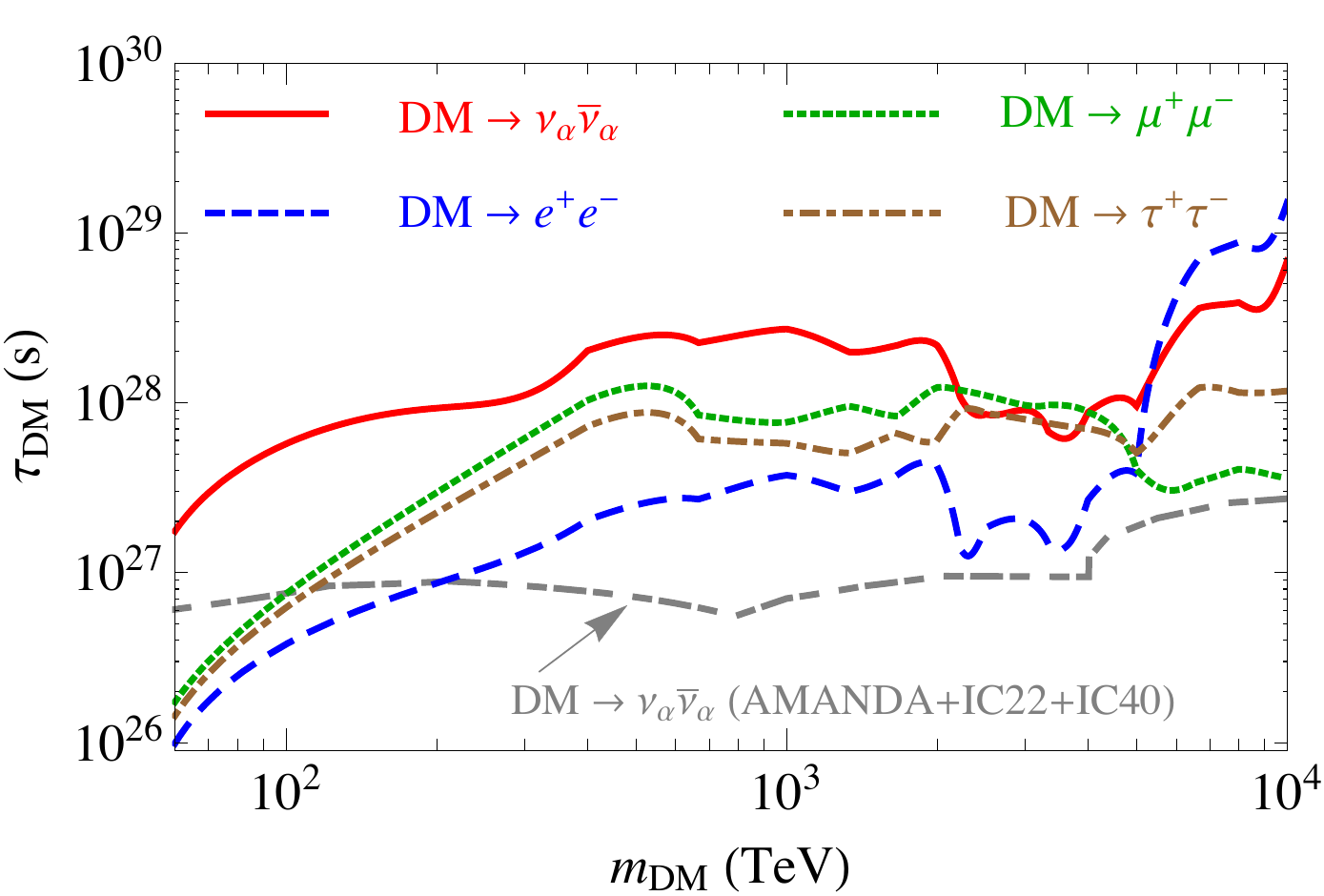}
\label{fig:lifetime-lep}
}
\subfloat[]{
\includegraphics[width=0.5\textwidth]{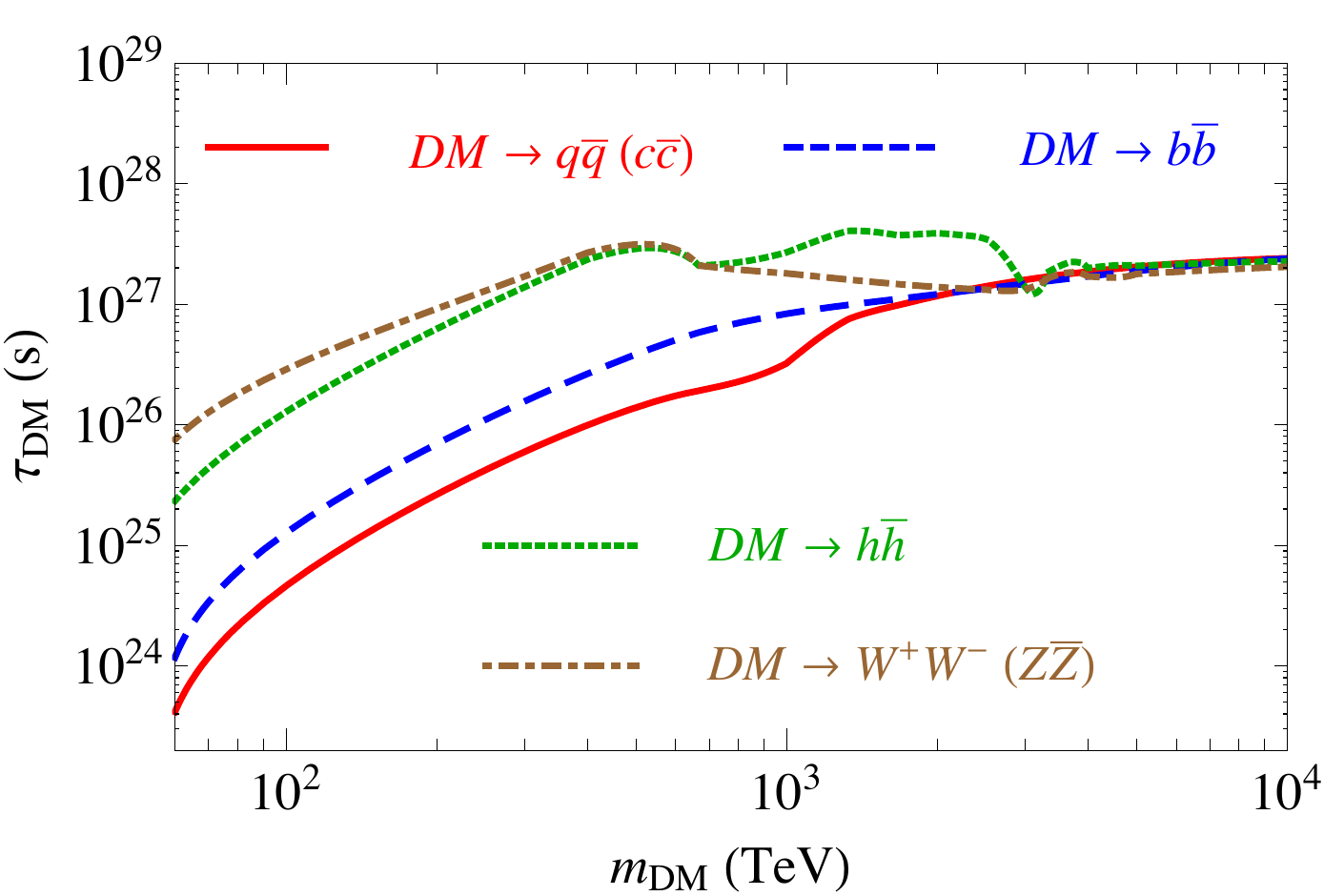}
\label{fig:lifetime-had}
}
\caption{\label{fig:lifetime} Lower bound on DM lifetime for various DM decay channels, at 90\% C.L., assuming an $E_\nu^{-2}$ astrophysical flux. } 
\end{figure}

\begin{figure}[t!]
\centering
\subfloat[]{
\includegraphics[width=0.5\textwidth]{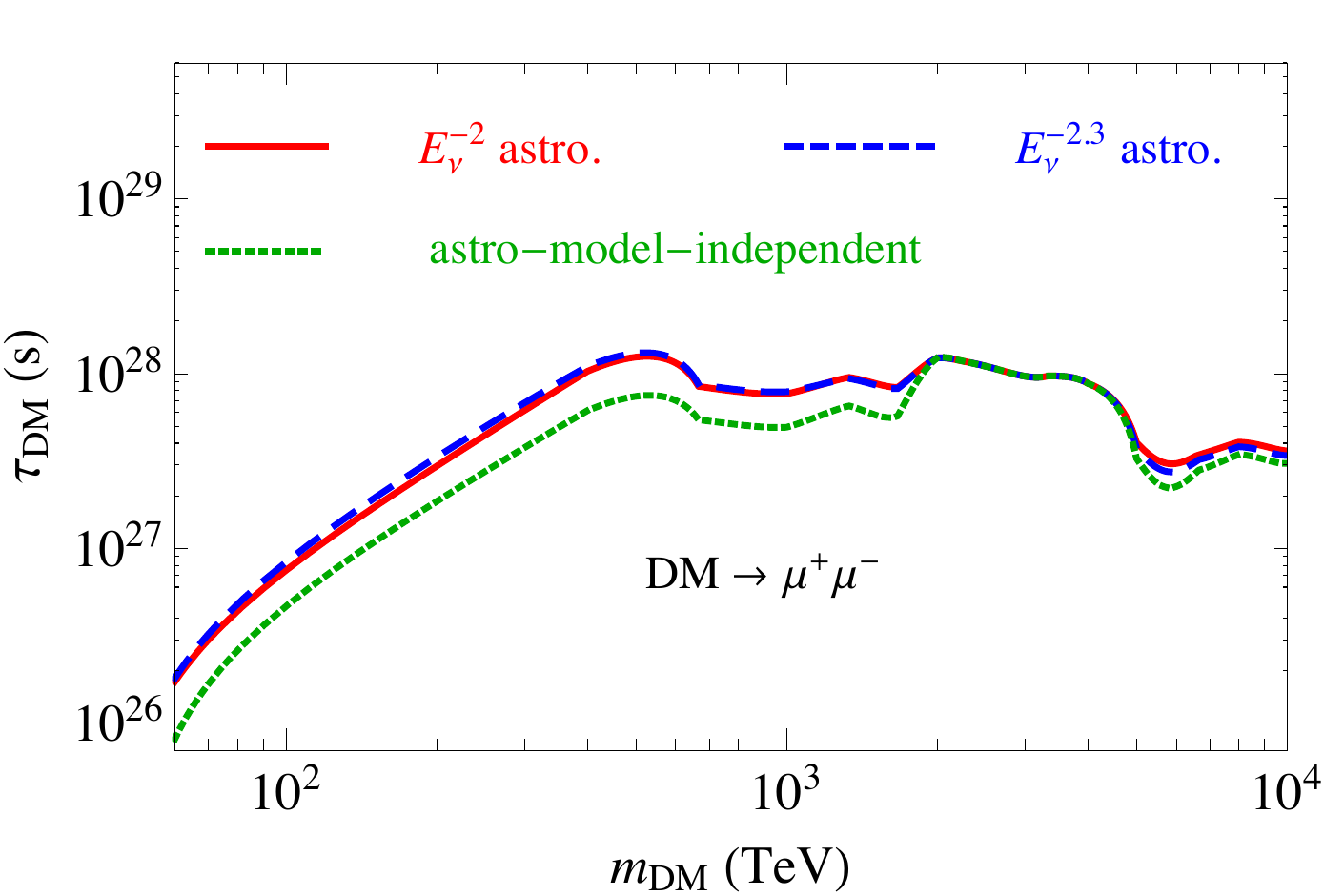}
\label{fig:lifetime-various}
}
\subfloat[]{
\includegraphics[width=0.5\textwidth]{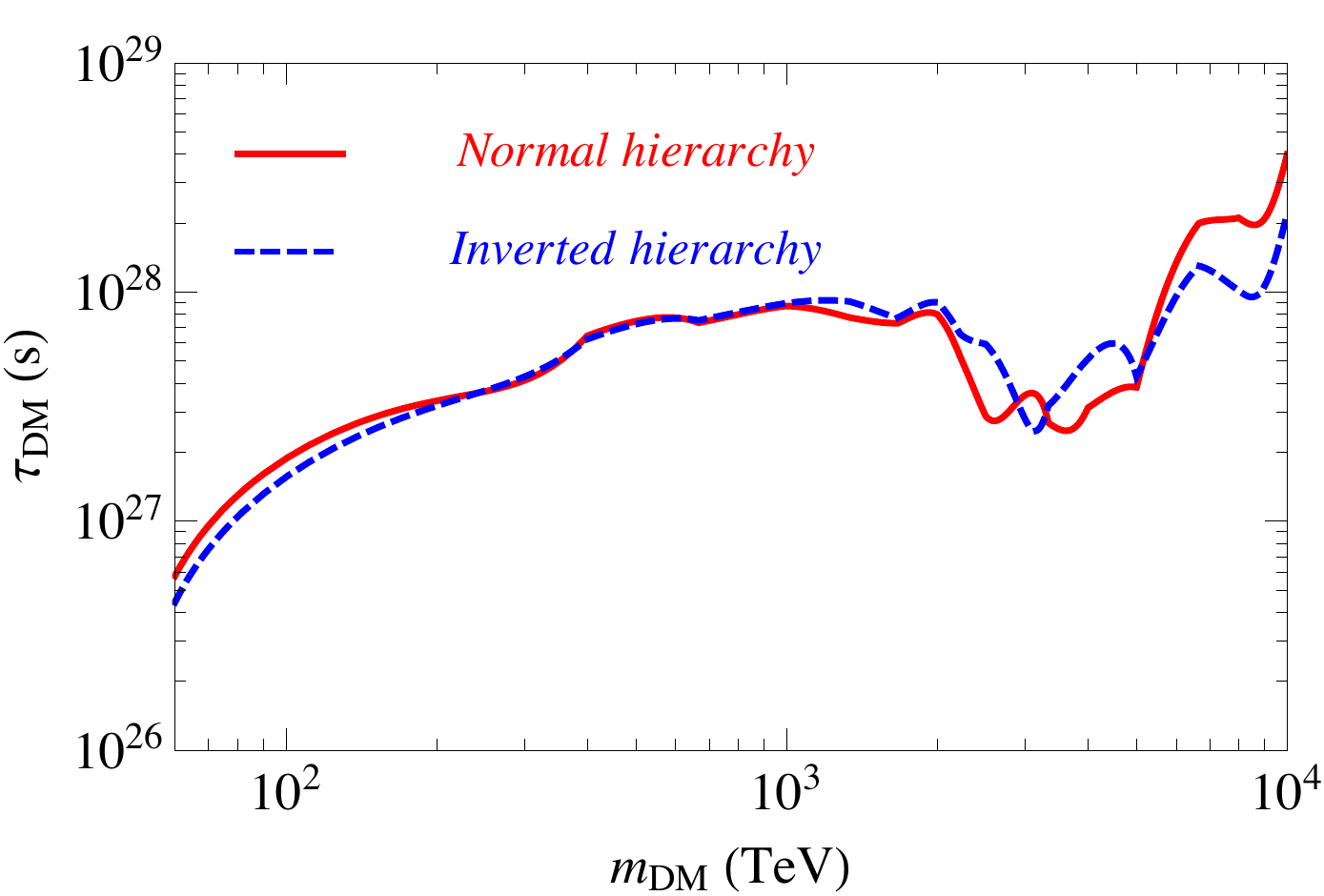}
\label{fig:limitNHIH}
}
\caption{\label{fig:NHIH} Panel (a): lower bound on DM lifetime for the model proposed in~\cite{Higaki:2014dwa}. The red solid line is for the NH model and the blue dashed line for IH. Panel (b): comparison of lower limit on $\tau_{\rm DM}$, for the channel ${\rm DM}\to\mu^+\mu^-$, for different assumptions for astrophysical flux: the red solid and blue dashed lines are for $E_\nu^{-2}$ and $E_\nu^{-2.3}$ astrophysical fluxes, respectively. The green dotted line is for the astro-model-independent case (i.e. using $N_{\rm limit}$ from the last column of Table~\ref{tab:numbers}). All the curves are at 90\% C.L..}
\end{figure}

Figure~\ref{fig:lifetime} shows the obtained lower limit on the DM lifetime for various DM decay channels: Figure~\ref{fig:lifetime-lep} for leptonic final states and Figure~\ref{fig:lifetime-had} for hadronic, gauge boson and Higgs final states. The final state $\nu_\alpha\bar{\nu}_\alpha$ in Figure~\ref{fig:lifetime-lep} means DM decaying to all the flavors of neutrinos with equal branching ratios. The gray dashed curve in Figure~\ref{fig:lifetime-lep} shows the lower limit on DM lifetime obtained before the recent IceCube data, taken from~\cite{Esmaili:2012us}, which should be compared with the red solid curve. As can be seen, by assuming an astrophysical origin for IceCube data, the lower limit on $\tau_{\rm DM}$ can be improved by about one order of magnitude. In both panels in Figure~\ref{fig:lifetime} we assumed a $E_\nu^{-2}$ astrophysical flux. However, to illustrate the dependence of the limits on the assumed astrophysical flux, in Figure~\ref{fig:lifetime-various} we plot the limit on $\tau_{\rm DM}$ for one channel, ${\rm DM}\to\mu^+\mu^-$, for both $E_\nu^{-2}$ and $E_\nu^{-2.3}$ astrophysical fluxes, as well as the model-independent limit described in eq.~(\ref{eq:L-in}). As it can be seen, the limits are the same for higher DM masses and only differ slightly at lower masses. Obviously, the model-independent limit is the most conservative one.      
 
Of course, in specific models of DM the lower limit on $\tau_{\rm DM}$ will depend on the combination of different branching ratios. As an example, Figure~\ref{fig:limitNHIH} shows the lower limit on $\tau_{\rm DM}$ for the model proposed in~\cite{Higaki:2014dwa}, with the red solid line for NH and the blue dashed line for IH. For any other model, the corresponding limit on lifetime can be derived from the curves in Figure~\ref{fig:lifetime} by appropriate scaling of the channels according to the branching ratios determined in the model. 

Note that independent constraints from diffuse gamma-rays yield bounds in the range of $10^{26}-10^{27}\,$s~\cite{Murase:2012xs,Cirelli:2012ut}, which makes the present bounds from neutrinos comparable if not better. In particular, for $\sim$~PeV masses and leptonic final states, these neutrino bounds are {\it stronger} by a factor of a few, which is consistent with the previous conclusions that the diffuse gamma-ray data are a factor of few below the sensitivity needed to probe the decaying DM models fitting the IceCube data. 

\section{Discussion and Conclusion}
\label{sec:conc}

The discovery of a flux of high energy neutrinos by the IceCube collaboration, originating from outside of the solar system, has ushered us into a new era for astroparticle physics. This observation opens a new window for exploring the high energy astrophysical sources, either Galactic or extragalactic ones, but also for constraining physics beyond the standard model. Since a consensus has still to be reached on the interpretation of the detected events, it is important to sharpen the diagnostic tools to probe different alternatives.  Here we have revisited the viability of a Dark Matter decay interpretation as the source of these events.

We have performed an extensive analysis of the angular distribution of the events.  With a variety of statistical tests (maximum likelihood, Kolmogorov-Smirnov and Anderson-Darling tests) we concluded that a mild preference (at or below the two sigma level) exists for a DM-like distribution with respect to a purely isotropic distribution foreseen for a conventional astrophysical flux of extragalactic origin. Qualitatively, the result does not depend on the type of statistical estimator used, nor from some of the approximations performed, and reflects an enhanced (albeit statistically not very significant) flux from the inner Galaxy region. We have also estimated that in a decade or so, IceCube should collect a sufficient statistics to perform a more constraining test, such that a DM distribution could be falsified at (close to) the three sigma level. Already at the present level, some aspects of our analyses are limited by the lack of knowledge of relevant experimental parameters. Thus, we definitely encourage the collaboration to provide a more stringent test of the DM hypothesis, together with the ongoing tests of the Galactic Plane component or of the compatibility with isotropy.

Concerning the energy spectrum, although we are not performing extended fits to the spectral shape, it is clear that DM models and a typical astrophysical model can provide comparably good fits and that the data have little discriminatory power at the moment. We substantiated this point by showing that even the simplest and most constrained effective field theory operator (dimension four) allowing for heavy DM decay in portal type models, is capable to provide a satisfactory fit of the data, notably at the highest energies (above a couple of hundred TeV). Needless to say, in more complicated models with more free parameters and extra decay channels the fit can be improved. Like for the directional signature, qualitatively the DM and astrophysical models do show clear spectral differences. The two classes of models typically differ: i) at $E_\nu\gg 2\,$PeV, where DM presents an abrupt cutoff, while astrophysical models are often associated to a very mild decrease (and actually a peak at the Glashow resonance energy of about 6.4 PeV); ii) in the intermediate energy range, say between 200 and 800 TeV, where a more or less prominent dip of events can be accommodated in DM models, but would be more difficult to explain in the astrophysical models.  It is worth mentioning, however, that this ``dip'' is not extreme: the difference in the event rate at these intermediate energies is not expected to exceed a factor of a few between the two cases. Probably, a reduction of the errors bar size by a similar level (so, naively at least a factor four higher statistics) is needed before this diagnostic tool becomes more meaningful. It is worth mentioning that the spectral shape at lower energy (tens of TeV) is affected by the atmospheric background as well---which may be underestimated---so the agreement with the spectrum at the highest energy is certainly a more stringent test.

Of course, it is possible and in fact likely that these neutrinos are not due to DM decay. Even assuming that the IceCube data originate from conventional astrophysical sources, however, we could derive bounds on decaying DM for various final states. We have derived the lower limit on heavy dark matter lifetime which shows improvements by up to an order of magnitude with respect to the existing constraints. That is by itself an interesting spin-off of the IceCube data for astroparticle physics. Also, in appendix~\ref{sec:ann} we showed that some non-trivial bounds can be obtained  on the annihilation cross section, $\langle \sigma v\rangle$, of WIMP-like DM with masses $\lesssim\,$100 TeV. For neutrino final states, these bounds are already stronger than the bounds from unitarity requirement. Definitely, by extending the threshold of the IceCube analysis to lower energies---it is encouraging that progress is being made in this sense~\cite{Aartsen:2014yta}--- these bounds could be made stronger.

Finally, we have commented on the complementary constraints on decaying DM from gamma-rays, which are not yet competitive with neutrino ones. We have shown that generally the available gamma-ray data are sensitive to PeV-scale DM lifetime $\sim 10^{26}-10^{27}$~s, which is one order of magnitude smaller than the lifetime required to interpret the IceCube neutrino data. The main obstacle to sharpening this sensitivity using the Fermi data is the yet unknown origin of the diffuse extragalactic background flux. However, this diagnostics can be improved by achieving a better understanding of this background and, to some extent, by resolving part of its sources. Another opportunity is the measurement of the photon fraction in high energy cosmic ray flux, as we mentioned in section~\ref{sec:gamma}. In particular, an ad hoc analysis (accounting for the peculiar anisotropy associated to the DM signal) of the photon fraction in cosmic ray data in the (0.1--1) PeV range can boost up current sensitivity. The IceCube collaboration itself, by using the IceTop facility, can contribute to progress in this direction.

Definitively, no matter what the nature of this newly discovered flux will turn out to be, these observations have contributed to open a new window to the universe and, at the same time, offered a serendipitous new probe of astroparticle physics.


\begin{acknowledgments}
At LAPTh, this activity was developed coherently with the research axes supported by the Labex grant ENIGMASS. For A.~E. this research was supported by the Munich Institute for Astro- and Particle Physics (MIAPP) of the DFG cluster of excellence ``Origin and Structure of the Universe''. P.~D.~S. would like to thank Seoul National University of Science and Technology for hospitality during the final phase of this work. We would like to thank Alejandro Ibarra for his useful comments on the manuscript. We thank E.~Resconi for clarifications on the flavor puzzle discussed in footnote~\ref{foot}.
\end{acknowledgments}

\appendix
\section{Bounds on the DM annihilation cross section}
\label{sec:ann}

The low energy part of the IceCube data can be used to update the limit on annihilation cross section $\langle \sigma v \rangle$ of DM. Since the flux of neutrinos from annihilating DM depends on the square of the DM halo density, effectively it drops very fast for DM masses $\gtrsim 100$~TeV (for which unitarity bounds become in general very severe, anyway). However, IceCube data extend down to $\sim30$~TeV and can be at least partly used to probe annihilating DM scenarios. 

The isotropic component of neutrino flux from annihilating DM has two contributions: 1) the residual isotropic flux from the Galactic halo; {\it i.e.}, the expected flux from the anti-GC direction; and 2) the cosmic flux from the annihilation of DM at all redshifts. In the following we calculate each of these contributions and, by confronting with the IceCube data, we derive the limit on $\langle \sigma v\rangle$ for some typical DM masses.   

\begin{itemize}

\item The residual isotropic neutrino flux from Galactic halo, from anti-GC direction, is given by
\begin{equation}
\frac{{\rm d}J_{\rm iso}^{\rm ann}}{{\rm d}E_\nu} = \frac{\langle \sigma v \rangle}{2} \frac{1}{4\pi m_{\rm DM}^2} \frac{{\rm d}N}{{\rm d}E_\nu} \left({\rm l.o.s.}\right)_{\rm anti-GC}~,
\end{equation}
where
\begin{equation}
\left({\rm l.o.s.}\right)_{\rm anti-GC} = \int_{0}^{\infty} \rho^2 [r(s,b=0,l=\pi)]~{\rm d}s~,
\end{equation}
is the line of sight integration at anti-GC direction. The average flux can be written as
\begin{equation}\label{eq:anti-gc}
\frac{{\rm d}J_{\rm iso}^{\rm ann}}{{\rm d}E_\nu} =  A_{\rm h}\; \frac{{\rm d}N}{{\rm d}E_\nu}~,
\end{equation}
where
\begin{equation}
A_{\rm h} =  6\times10^{-13} \left( \frac{\langle \sigma v \rangle}{10^{-22}~{\rm cm}^3~{\rm s}^{-1}} \right) \left( \frac{100~{\rm TeV}}{m_{\rm DM}} \right)^2 ({\rm cm}^2~{\rm s}~{\rm sr})^{-1}~.
\end{equation}
For our analysis we consider the three lowest energy bins of IceCube data (bins $\#1$, $\#2$ and $\#3$ in Table~\ref{tab:numbers}). The expected number of events in each bin can be calculated by convoluting the flux in eq.~(\ref{eq:anti-gc}) with the effective area of IceCube. Similar to the limit on DM lifetime in section~\ref{sec:lifetime}, the limit on $\langle\sigma v\rangle$ can be derived by the requirement that in all the bins the expected number of events from annihilating DM should be smaller than $N_{\rm limit}$, where here we use the values obtained by assuming an $E_\nu^{-2}$ astrophysical flux (see Table~\ref{tab:numbers}). The obtained limits on $\langle \sigma v \rangle$ for three values of DM mass ($m_{\rm DM}=30$, 50 and 100~TeV) are given in Table~\ref{tab:anti-gc}.

\begin{table}[t]
\caption{The upper limit on $\langle \sigma v \rangle$ (at 90\% C.L.) for various channels of DM annihilation and three masses of DM, obtained from the residual isotropic Galactic halo flux at anti-GC direction (see eq.~(\ref{eq:anti-gc})). The unit of $\langle \sigma v \rangle$ is $10^{-22}~{\rm cm}^3~{\rm s}^{-1}$.}
\begin{center}
\begin{tabular}{|c|c|c|c|}
\hline
\backslashbox{{\footnotesize DM + DM} $\to$}{$m_{\rm DM}$} & 100~TeV & 50~TeV & 30~TeV \\
\hline 
$\nu_\alpha\overline{\nu}_\alpha$ & 1.92 & 1.4 & 2.71 \\
\hline 
\parbox[c]{13pt}{$q\overline{q}$} & 632 & 1,767 & 10,653 \\
\hline 
\noindent\parbox[c]{13pt}{$b\overline{b}$} & 244 & 643 & 3,770 \\
\hline 
\noindent\parbox[c]{13pt}{$c\overline{c}$} & 771 & 2,129 & 12,025 \\
\hline 
$e^+e^-$ & 20 & 20.7 & 46.7 \\
\hline 
$\mu^+\mu^-$ & 5.5 & 10.2 & 26 \\
\hline 
$\tau^+\tau^-$ & 7.2 & 12.3 & 31.5 \\
\hline 
\noindent\parbox[c]{16pt}{$h\overline{h}$} & 25.8 & 59 & 191 \\
\hline 
\noindent\parbox[c]{23pt}{$Z\bar{Z}$} & 14.7 & 20 & 44.5 \\
\hline 
$W^+W^-$ & 17.6 & 26 & 59.5 \\
\hline 
\end{tabular}
\end{center}
\label{tab:anti-gc}
\end{table}

\item The limits in Table~\ref{tab:anti-gc} are obtained by considering merely the Galactic halo contribution; in this sense, they are very conservative. However, a second contribution to isotropic neutrino flux is the cosmic flux from the annihilation of DM at all the redshifts. The cosmic flux is given by
\begin{equation}\label{eq:cosmic}
\frac{{\rm d}J^{\rm ann}_{\rm cos}}{{\rm d}E_\nu} = \frac{\langle \sigma v \rangle}{2} \frac{\Omega_{\rm DM}^2 \rho_c^2}{4\pi m_{\rm DM}^2} \frac{c}{H_0} \int_{0}^\infty \frac{(1+z)^3\zeta(z){\rm d}z}{\sqrt{\Omega_m (1+z)^3 + \Omega_\Lambda}} \frac{{\rm d}N}{{\rm d}E_\nu} \left[\left(1+z\right)E_\nu\right]~,
\end{equation}
where $\zeta(z)$ takes into account the DM clustering which depends on the redshift. There are several calculation of $\zeta(z)$ in the literature (see for instance~\cite{Sefusatti:2014vha,Ullio:2002pj,Bergstrom:2001jj,Taylor:2002zd,Maccio':2008xb,Neto:2007vq}). Although most of the estimations of $\zeta(z)$ match at large $z$ (where contribute to cosmic flux negligibly), at small $z$ values, where the significant part of flux comes from, large discrepancies exist. A recent overview and re-evaluation of the uncertainties in $\zeta(z)$ can be found in~\cite{Sefusatti:2014vha}, which represents a follow-up and improvement of the method proposed in~\cite{Serpico:2011in}. For our analysis, we take two extreme cases of $\zeta(z)$, from Figure~5 of~\cite{Sefusatti:2014vha}, obtained from the result of Millennium Simulation II~\cite{BoylanKolchin:2009nc}. Table~\ref{tab:cosmic} reports the obtained limits on $\langle\sigma v\rangle$, considering both the fluxes in eqs.~(\ref{eq:anti-gc}) and (\ref{eq:cosmic}), for various channels of DM annihilation. The reported limits correspond to min$\div$max value used for $\zeta(z)$, as we described.    

\begin{table}[h]
\caption{The upper limit on $\langle \sigma v \rangle$ (at 90\% C.L.) for various channels of DM annihilation and three masses of DM, obtained from considering both the residual isotropic Galactic halo flux at anti-GC direction, eq.~(\ref{eq:anti-gc}), and cosmic flux, eq.~(\ref{eq:cosmic}). The unit of $\langle \sigma v \rangle$ is $10^{-22}~{\rm cm}^3~{\rm s}^{-1}$. In each cell, $\#\div\#$ correspond to minimum$\div$maximum values used for $\zeta(z)$.}
\begin{center}
\begin{tabular}{|c|c|c|c|}
\hline
\backslashbox{{\footnotesize DM + DM} $\to$}{$m_{\rm DM}$} & 100~TeV & 50~TeV & 30~TeV \\
\hline 
$\nu_\alpha\overline{\nu}_\alpha$ & 1.39~$\div$~0.22 & 1.21~$\div$~0.36 & 2.44~$\div$~0.88 \\
\hline 
\parbox[c]{13pt}{$q\overline{q}$} & 489~$\div$~84.5 & 1427~$\div$~299 & 9934~$\div$~4603 \\
\hline 
\noindent\parbox[c]{13pt}{$b\overline{b}$} & 185~$\div$~30.4 & 517~$\div$~106 & 3514~$\div$~1621 \\
\hline 
\noindent\parbox[c]{13pt}{$c\overline{c}$} & 592~$\div$~100 & 1708~$\div$~348 & 11218~$\div$~5215 \\
\hline 
$e^+e^-$ & 14.7~$\div$~2.38 & 17.8~$\div$~5.06 & 41.3~$\div$~14.2 \\
\hline 
$\mu^+\mu^-$ & 4.47~$\div$~0.65 & 9.06~$\div$~1.6 & 23.7~$\div$~9.23 \\
\hline 
$\tau^+\tau^-$ & 5.84~$\div$~0.93 & 10.9~$\div$~2.3 & 28.5~$\div$~10.8 \\
\hline 
\noindent\parbox[c]{16pt}{$h\overline{h}$} & 21.2~$\div$~3.36 & 53.4~$\div$~9.49 & 177~$\div$~76.5 \\
\hline 
\noindent\parbox[c]{23pt}{$Z\bar{Z}$} & 11.9~$\div$~2.05 & 18.1~$\div$~4.09 & 40.7~$\div$~16.3 \\
\hline 
$W^+W^-$ & 14.4~$\div$~2.4 & 23.7~$\div$~4.96 & 54.5~$\div$~22.3 \\
\hline 
\end{tabular}
\end{center}
\label{tab:cosmic}
\end{table}

As can be seen, for the minimum values used for $\zeta(z)$, the limit on $\langle \sigma v\rangle$ in Table~\ref{tab:cosmic} is close to the one obtained by considering just the Galactic halo contribution (see Table~\ref{tab:anti-gc}) since in this case the residual isotropic Galactic halo contribution is the dominant flux. On the other hand, for the maximum values used for $\zeta(z)$, the cosmic flux dominates and the limits on $\langle\sigma v\rangle$ are stronger by almost an order of magnitude. For the neutrino final state case, these limits are more stringent than the ones imposed by unitarity constraints (see for instance Fig.~2 in~\cite{Kachelriess:2007aj}) and might thus be of some phenomenological interest.

\end{itemize}
 \newpage

\end{document}